\documentclass[useAMS,usenatbib,usegraphicx]{mn2e}

\usepackage{units,amssymb,url}
\def\aap{A\&A}                
\def\msol{M$_\odot$yr$^{-1}$}

\title[The chemistry of extragalactic carbon stars]{The chemistry of extragalactic carbon stars}
\author[Paul~M.~Woods et al.]
{Paul M.~Woods$^{1,2}$\thanks{E-mail: pmw@star.ucl.ac.uk},
C.~Walsh$^3$,
M.~A.~Cordiner$^4$,
F.~Kemper$^{5}$
\\$^{1}${Department of Physics \& Astronomy, University College London, Gower Street, London, WC1E 6BT, UK}
\\$^{2}${Jodrell Bank Centre for Astrophysics, Alan Turing Building, School of Physics and Astronomy, The University of Manchester,}
\\\phantom{$^2$}{Oxford Road, Manchester, M13 9PL, UK}
\\$^{3}${Astrophysics Research Centre, School of Mathematics \& Physics, 
Queen's University Belfast, University Road, Belfast, BT7 1NN, UK}
\\$^{4}${Astrochemistry Laboratory and The Goddard Center for Astrobiology, 
NASA Goddard Space Flight Center, Code 691,}
\\\phantom{$^4$}{8800 Greenbelt Road, Greenbelt, MD 20771, USA}
\\$^{5}${Academia Sinica Institute of Astronomy \& Astrophysics, PO Box
23-141, Taipei 10617, Taiwan}
}
\begin{document}

\date{}

\pagerange{\pageref{firstpage}--\pageref{lastpage}} \pubyear{2012}

\maketitle

\label{firstpage}

\begin{abstract}
  Prompted by the ongoing interest in Spitzer Infrared Spectrometer
  spectra of carbon stars in the Large Magellanic Cloud, we have
  investigated the circumstellar chemistry of carbon stars in
  low-metallicity environments. Consistent with observations, our
  models show that acetylene is particularly abundant in the inner
  regions of low metallicity carbon-rich AGB stars -- more abundant
  than carbon monoxide. As a consequence, larger hydrocarbons have
  higher abundances at the metallicities of the Magellanic Clouds than
  in stars with solar metallicity. We also find the oxygen and
  nitrogen chemistry is suppressed at lower metallicity, as
  expected. Finally, we calculate molecular line emission from carbon
  stars in the Large and Small Magellanic Cloud and find that several
  molecules should be readily detectable with the Atacama Large
  Millimeter Array at Full Science operations.
\end{abstract}


\section{Introduction}

The advent of the Atacama Large Millimeter Array (ALMA) and other new
(sub-)millimetre telescopes, with their unprecendented spatial
resolution and sensitivity, will allow the observation of giant stars
in other galaxies in similar detail to that achieved for Galactic
objects. These advanced capabilities prompt investigation into the
nature of these extragalactic stars, with one of the most interesting
aspects being the study of the effect of sub-solar metallicities on
circumstellar chemistry and dust composition.

Recent studies in the infrared using the Spitzer Space Telescope and
ground-based instruments have highlighted the deep molecular
absorptions of, primarily, acetylene in the spectra of evolved carbon
stars, in the Magellanic Clouds
\citep[e.g.,][etc.]{vlo99a,mat02,mat05,vlo06,spe06,zij06,slo06,lag07,lei08,vlo08,woo11}. These
absorption features are in general deeper than those seen in Galactic
stars, and this implies that there is a difference between the
chemistry of Magellanic circumstellar envelopes (CSEs) and Galactic
carbon stars, which are comparatively well-studied.

The Magellanic Clouds (MCs) are nearby dwarf galaxies with sub-solar
metallicities. The Large Magellanic Cloud (LMC), at a distance of
$\sim$50\,kpc \citep{sch08}, has an average metallicity of
around 50\% solar \citep*{duf82,wes97}. The Small Magellanic Cloud
(SMC), at a slightly larger distance of 66\,kpc \citep{sze09}, has an
average metallicity of 20\% solar \citep{duf82,wes97}. The
effect of this low metallicity regime on dust composition and galactic
dust budgets has been studied observationally by many authors
\citep[e.g.,][]{zij06,lag07,vlo08,mat09}. However, the effects on
circumstellar chemistry have not been, as yet, studied in any detail,
and our aim here is to pioneer in the field with this work.

The chemical modelling of Galactic AGB stars was first attempted 35
years ago. Initial attempts focused on simple physical models and
chemistry appropriate for oxygen-rich ($n(\mathrm{O})>n(\mathrm{C}$))
circumstellar environments \citep{gol76,sca80,jur81}. Physical models
have largely remained simple \citep[with some notable exceptions,
  e.g.,][]{cor09} whereas the chemical modelling has moved to focus on
carbon-rich ($n(\mathrm{C})>n(\mathrm{O}$)) circumstellar chemistry
since it shows a wider variety of molecules
\citep[e.g.,][]{hug82}. Progress in chemical modelling is driven in
part by the desire to explain observed abundances of newly-detected
molecules in the most accessible carbon star, IRC+10216, for instance,
in the addition of long carbon-chain molecules \citep{mil00} or anion
species \citep{mil07}. In this case, advances in technology have
driven us to investigate carbon-rich circumstellar chemistry in
previously challenging locations.

In this paper, we present the results of modelling the circumstellar
chemistry around carbon-rich AGB stars at sub-solar metallicities. We
focus on three fiducial models, at metallicities and with physical
conditions appropriate for the Galaxy, the LMC and the SMC. We
initially assume solar metallicity ($Z$=0.02) for Galactic carbon
stars, and average interstellar metallicities for LMC ($Z$=0.008) and
SMC ($Z$=0.004) carbon stars. In \S2 we describe our adopted physical
model and discuss the different physical and chemical
considerations, and also discuss sources of uncertainties in our
models in \S2.6. In \S3 we describe how we calculate the chemical
evolution of the circumstellar envelope. We show our results in \S4,
preceding a discussion in \S5 (including our calculations of molecular
line emission which may be observable with ALMA) and finally, in \S6,
we draw our conclusions regarding the chemistry of low-metallicity
carbon stars.





\section{Chemical and physical considerations at low metallicity}

Nucleosynthetic products dredged (via convection) from the stellar
interior are mixed to the surface of the star, where material is
accelerated to the terminal velocity of the stellar wind within a
radius of 20\,R$_\star$\ \citep{kea88} and passed into the CSE. The
gas, which is mainly molecular hydrogen, is well-mixed with dust
grains. We assume a spherical geometry where the gas has a 1/$R^2$\
density distribution, and a temperature profile which follows
$$ T(r) = \mathrm{max}[150(R/\mathrm{R_0})^{-0.79};
  10]\mathrm{\,K.}\quad(R \geq \mathrm{R_0} =
  5\times10^{15}\:\mathrm{cm})$$ \citep[e.g.,][]{mil00,mil94}. The CSE
  is irradiated by the interstellar radiation field (ISRF), but not by
  UV photons from the star itself, which are quenched in the stellar
  atmosphere. Extinction in the CSE is calculated according to the
  approach of \citet{jur81}, assuming interstellar-type grains, and we
  treat CO self-shielding according to \citet{mam88}. We model the
  carbon-rich chemistry in the circumstellar region between
  $\sim$200--100\,000\,R$_\star$\ (0.005--3$\times$10$^{18}$\,cm).

In the subsequent sections, we consider the chemical and further
physical ingredients of our model. Observationally, very little is
known about the chemistry of extragalactic carbon stars, and so we
mainly discuss the chemistry of Galactic carbon stars, using the
nucleosynthesis models of \citet{kar10} to adjust for lower
metallicities (see \S\ref{sec:initabunds}). Physical parameters of LMC
and SMC carbon stars are more well constrained observationally, and we
summarise those aspects in \S\ref{sec:mlr}--\ref{sec:isrf}.

\subsection{Initial chemical abundances}
\label{sec:initabunds}

\begin{table*}
\begin{minipage}{165mm}
\caption{Initial model fractional abundances (w.r.t. H$_2$)}
\label{tab:initabunds}
\begin{tabular}{@{}lccccc}
\hline
Species & Observed Galactic inner & TE abundance   &  Adopted initial      & Adopted initial & Adopted initial \\
        & envelope abundance      & (Galactic)     &  abundance (Galactic) & abundance (LMC) & abundance (SMC) \\
\hline
CO         &$^{a}$6$\times$10$^{-4}$ & 4.4$\times$10$^{-4}$ & 4.4$\times$10$^{-4}$ & 1.7$\times$10$^{-4}$ & 8.9$\times$10$^{-5}$\\
N$_2$      &$^{b}$ ---               & 5.4$\times$10$^{-5}$ & 5.4$\times$10$^{-5}$ & 1.2$\times$10$^{-5}$ & 3.9$\times$10$^{-6}$\\
C$_2$H$_2$ &$^{c}$8$\times$10$^{-5}$ & 3.3$\times$10$^{-5}$ & 3.3$\times$10$^{-5}$ & 2.8$\times$10$^{-4}$ & 3.1$\times$10$^{-4}$\\
HCN        &$^{d}$2$\times$10$^{-5}$ & 2.4$\times$10$^{-5}$ & 2.4$\times$10$^{-5}$ & 3.3$\times$10$^{-5}$ & 2.0$\times$10$^{-5}$\\
SiS        &$^{e}$3$\times$10$^{-6}$ & 3.7$\times$10$^{-6}$ & 6.9$\times$10$^{-6}$ & 2.1$\times$10$^{-6}$ & 1.0$\times$10$^{-6}$\\
CS         &$^{c}$4$\times$10$^{-6}$ & 4.7$\times$10$^{-6}$ & 2.4$\times$10$^{-6}$ & 1.6$\times$10$^{-6}$ & 8.0$\times$10$^{-7}$\\
NH$_3$     &$^{f}$2$\times$10$^{-6}$ & 3.3$\times$10$^{-11}$ & 2.0$\times$10$^{-6}$ & 8.0$\times$10$^{-7}$ & 4.0$\times$10$^{-7}$\\
CH$_4$     &$^{c}$2$\times$10$^{-6}$ & 6.1$\times$10$^{-9}$ & 2.0$\times$10$^{-6}$ & 8.0$\times$10$^{-7}$ & 4.0$\times$10$^{-7}$ \\
SiO        &$^{g}$1$\times$10$^{-6}$ & 2.0$\times$10$^{-8}$ & 1.7$\times$10$^{-6}$ & 6.8$\times$10$^{-7}$ & 3.6$\times$10$^{-7}$\\
SiH$_4$    &$^{c}$2$\times$10$^{-7}$ & 8.8$\times$10$^{-15}$ & 2.0$\times$10$^{-7}$ & 8.0$\times$10$^{-8}$ & 4.0$\times$10$^{-8}$\\
\hline
\end{tabular}
$^{a}${\citet{kwa82}}\quad $^{b}${Not observed due to lack of
  permanent dipole. 2$\times$10$^{-4}$ is usually assumed in models
  \citep[e.g.,][]{cor09}}, similar to the elemental abundance\quad
$^{c}${\citet{kea93,cer99}}\quad $^{d}${\citet{sch07a,cer99}}\quad
$^{e}${\citet{sch07b,dec10}}\quad $^{f}${\citet{has06,mon00,kea93}}\quad
$^{g}${\citet{sch06,kea93}}
\end{minipage}
\end{table*}

As circumstellar material cools during the expansion of the stellar
  wind, it is energetically favourable for atomic elements to form
  molecules in the photosphere of the star ($R\lse5$\,R$_\star$),
  where the chemistry is in thermal equilibrium (TE) due to the high
  temperatures and densities. For molecules of high stability, that is
  to say, those which contain strong internal bonds, the abundances
  amounted in this region of the star are carried through into the
  non-TE outer envelope ($R\gse 100$\,R$_\star$) without significant
  change. In carbon-rich environments such molecules include CO,
  N$_2$, HCN and C$_2$H$_2$ \citep[e.g.,][]{tsu73,mil08}.

A further group of molecules are also observed at high abundances at
the inner edge of the outer envelope in Galactic stars. This subset is
generally formed at low abundance in the TE photosphere of the star,
but attains a higher abundance in the inner envelope where dust is
formed and then accelerated ($5\lse R \lse 100$\,R$_\star$) due to
non-equilibrium processes. They comprise CS, SiO, SiS, NH$_3$, SiH$_4$
and CH$_4$\ \citep[e.g.,][]{che06}. These two subsets of molecules are
often called ``parent species''.


\begin{figure}
\includegraphics[width=84mm]{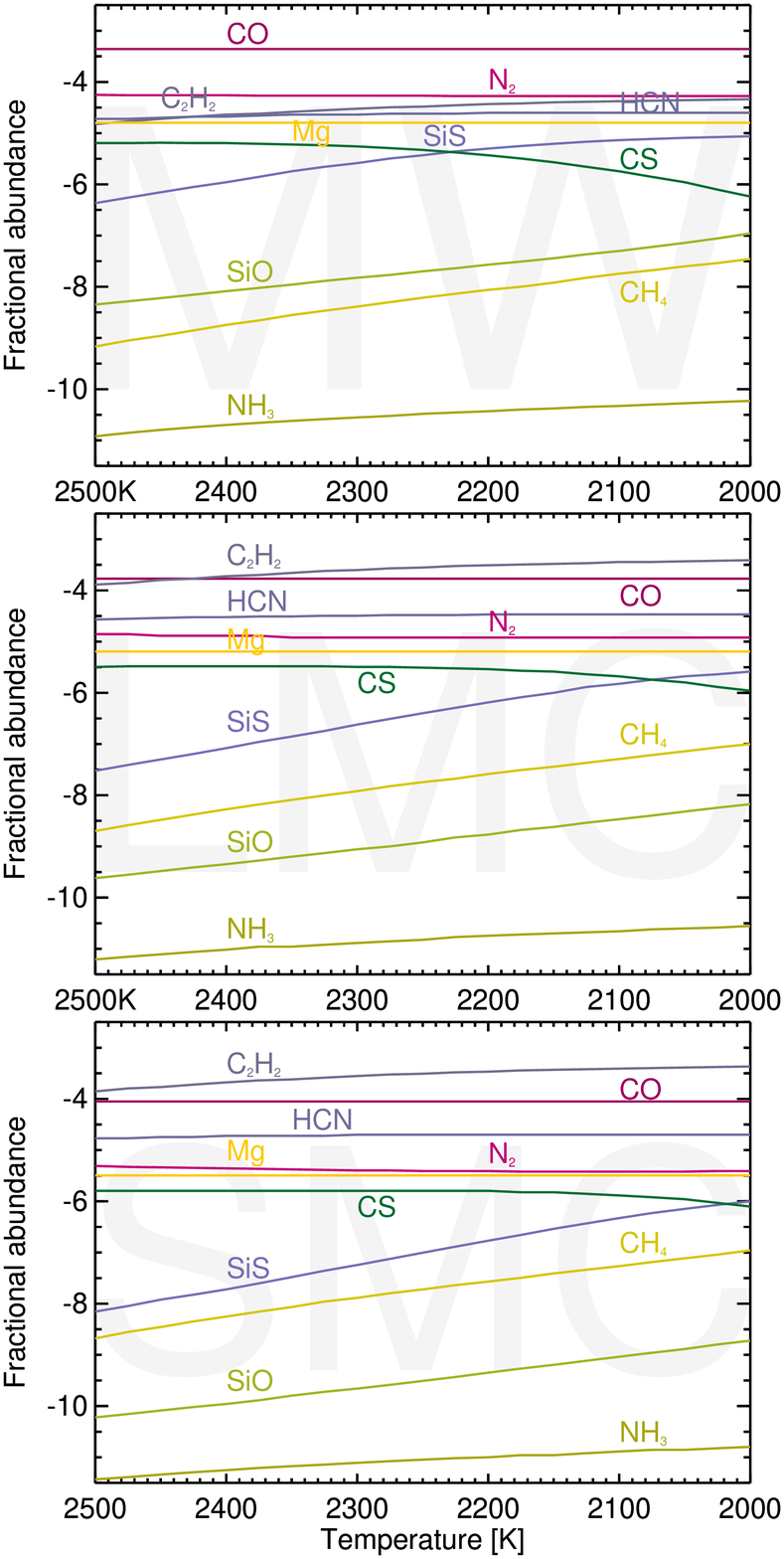}
\caption{Fractional abundances (log scale) at thermal equilibrium for
  MW, LMC and SMC metallicities, at a pressure of
  P$\sim$10$^{-3}$\,atm. We utilise values at 2\,250\,K in the
  subsequent modelling.}
\label{fig:te}
\end{figure}

To obtain accurate initial abundances for these ten parent species
when modelling the chemistry in the outer envelope ($R\gse
100$\,R$_\star$) at different metallicities, one must employ different
techniques for the two subsets of parents. For the high stability
parent species we can use a TE model to calculate the relevant
data. This approach has been used previously to good effect
\citep[e.g.,][]{tsu73,sha90,mar00}. For the remaining parent species
other physical factors must be taken into consideration, such as
depletion through dust formation \citep[SiS, SiO;][]{bie89,sch06},
pulsation-driven shocks \citep[SiS, SiO, CS;][]{wil98} and gas-grain
interaction (e.g., hydrogenation leading to NH$_3$, SiH$_4$ and
CH$_4$). Modelling these complexities in detail, which indeed
themselves have been the subject of much investigation, is too
advanced for this initial study and thus we must use other arguments,
discussed below.

\subsubsection{TE molecules (CO, N$_2$, C$_2$H$_2$, HCN)}

The TE model used is similar to that detailed by \citet{sha90} and
\citet{mar00}\footnote{Available on request to the authors.}, which
work by minimising the Gibbs free energy of the thermodynamic
system. We adopt a temperature of 2\,250\,K and a pressure of
1.033$\times$10$^{-3}$\,atm, appropriate for the photosphere of a
carbon-rich AGB star \citep{ive96,mar00}. Results of the TE
calculations for the range of temperatures 2\,500--2\,000\,K are
displayed in Fig.~\ref{fig:te}; those species for which TE is
appropriate are largely invariant across the temperature
range. Elemental abundances for AGB stars at different metallicities
are taken from the nucleosynthesis calculations of \citet{kar10} for a
3\,M$_\odot$ star. This mass was chosen because it was the
lowest--mass model for which the star became carbon-rich at the three
metallicites $Z$=0.02 (Galaxy), $Z$=0.008 (LMC) and $Z$=0.004
(SMC). There are incongruities with these models, in that it is clear
from observational evidence that carbon stars in the Milky Way can
form at masses as low as $\sim$1.5\,M$_\odot$ \citep{wal98}. These
modelling issues are acknowledged and being addressed \citep{kar11},
and we continue to use the results since they are a consistent set of
readily-available abundances. Average stellar yields were used as
inputs for the TE model and the resulting fractional abundances with
respect to H$_2$ can be found in Table~\ref{tab:initabunds} for CO,
N$_2$, C$_2$H$_2$ and HCN. This method (using the results of
nucleosynthesis calculations), is preferable to using the elemental
abundances of the Magellanic Clouds in general, since much of the
circumstellar chemistry depends on the amount of carbon generated by
the star in the AGB phase \citep{mats08}.  In the next section, we
describe our method of determining input abundances for those
molecules for which TE does not apply.

\subsubsection{Non-TE molecules (CS, SiO, SiS, NH$_3$, SiH$_4$, CH$_4$)}

The observed distribution of SiO peaks within a few stellar radii in
the archetypal carbon star, IRC+10216 \citep{sch06,kea93}. SiO is seen
at high fractional abundances, n(SiO)/n(H$_2$)$\sim$10$^{-6}$, whereas
TE models underpredict the abundance by around a factor of 30. Its
formation is thought to be due to circumstellar shocks, and hence its
abundance is dependent on that of atomic oxygen, since in the shocked
material the reaction:
\begin{equation}
\rm Si\ + OH\ \longrightarrow\ SiO\ + H
\end{equation} 
dominates \citep*{wil98,har80}. As such, we choose to set the
fractional abundance of SiO to 10$^{-3}$n(O), in line with that
observed in IRC+10216.

SiS is the main repository for both silicon and sulphur in the gas
phase, and forms at a high abundance in TE models. Observations
indicate that it is rapidly depleted as gas flows through the
intermediate envelope, presumably due to the adsorption of the
molecule onto grain surfaces \citep{bie89,boy94}. Thus we assume that
the initial abundance of SiS is equal to the elemental abundance of
sulphur minus the fraction of CS (discussed below). For IRC+10216 this
premise agrees well with observations \citep{bie93,luc95,sch06,dec10}.

CS is slightly overproduced in TE models compared to observations
\citep{kea93} and work by \citet{wil98} has shown that CS is destroyed
by pulsation-driven shocks by a considerable amount, the extent of
which depends on the assumed shock speed. It then slowly reforms in
the outflowing wind. We assume slow shock speeds (10\,km\,s$^{-1}$)
and the same shock-destruction efficiency as \citet{wil98}, and adopt
an abundance of CS a factor of 2 less than its TE abundance.

Emission from the hydrogenated species NH$_3$, CH$_4$ and SiH$_4$
originates in the dust-formation zone and inner envelope around
IRC+10216 \citep{kea93}. Although surprising because of the high
temperature of the dust in these regions, these molecules are most
likely formed via hydrogen-addition reactions on the surface of dust
grains. The very short time scale before ejection from the grain
surface would indicate that hydrogen is the dominant reaction partner
for adsorbed atoms. Typically these molecules are of low abundance in
both TE models \citep[e.g.,][]{tsu64}, and non-TE models \citep[][and
  others]{wil98}, adding further weight to grain-surface hydrogenation
theories. Observations of warm NH$_3$ in IRC+10216 have been used to
derive fractional abundances of
0.2--2.0$\times$10$^{-6}$\ \citep{kea93,mon00,has06}. We assume that
the abundance of NH$_3$ at other metallicites will scale with the dust
surface area. For the LMC and SMC this quantity, and the associated
gas-to-dust ratio and grain-size distribution, are \emph{highly}
uncertain, although work is currently ongoing to determine more
definite values. Assuming that the grain-size distribution in LMC and
SMC carbon stars is similar to that in Galactic carbon stars, and that
the gas-to-dust ratio scales with metallicity, we will adopt the upper
observed fractional abundance of NH$_3$ for Galactic carbon stars, and
scale according to metallicity for the LMC and SMC carbon stars. We
apply a similar reasoning for scaling the initial fractional
abundances of CH$_4$ and SiH$_4$ with metallicity.

Adopting the initial abundances detailed in Table~\ref{tab:initabunds}
means that only small amounts of nitrogen, oxygen and sulphur are
available for incorporation into other molecules and dust. However, in
the Galactic case, some 5\% of the elemental carbon remains for
incorporation into molecules or carbonaceous dust. At LMC and SMC
metallicities, the percentage of free carbon rises to 28\% and 32\%,
respectively, which is in reasonable agreement with \citet{fer06} for
an evolved star. These authors also found that at most 9\% of
elemental silicon condensed into dust. Our initial abundances mean
that 52--60\% of the elemental silicon is available for dust or dust
seed formation.


In Table~\ref{tab:COratios} we have compiled a comparison of C/O
ratios and the carbon excesses
(log($\epsilon_\mathrm{C}-\epsilon_\mathrm{O}$)) for the three metallicity
regimes. C/O ratios increase drastically in the nucleosynthesis
calculations of \citet{kar10}, largely due to the decrease in
elemental oxygen. The carbon excess is larger at low metallicity, an
effect which has been observed \citep[e.g.,][]{wah06}.

\begin{table}
  \caption{C/O ratios, elemental carbon and oxygen abundances ($\epsilon_\mathrm{C}$,
    $\epsilon_\mathrm{O}$), and carbon excesses in the three metallicity environments.}
\label{tab:COratios}
\begin{tabular}{@{}lcccc}
\hline
Environment & $\epsilon_\mathrm{C}$ & $\epsilon_\mathrm{O}$ & C/O & C$-$O\\
\hline
Milky Way & 9.05 & 8.94 & \phantom{1}1.3 & 2.5$\times$10$^{-4}$ \\
LMC       & 9.33 & 8.53 & \phantom{1}6.3 & 1.8$\times$10$^{-3}$  \\
SMC       & 9.33 & 8.25 & 12.0           & 2.0$\times$10$^{-3}$  \\
\hline
\end{tabular}
\end{table}

\subsection{Mass-loss rate}
\label{sec:mlr}

Mass-loss rates for AGB stars are generally measured using
observations of the CO envelope and then assuming a
CO-to-H$_2$\ scaling ratio (the so-called \textquoteleft
X-factor\textquoteright, $X_\mathrm{CO}$). Alternatively, the near-
and mid-infrared spectral energy distribution can be fitted with a
radiative-transfer model to obtain a dust mass-loss rate, and then by
assuming a gas-to-dust ratio, the total (gas+dust) mass-loss rate can
be calculated. The former method has been routinely used for Galactic
stars, but extragalactic stars are in general too faint. The
availability of high-sensitivity data from, for example, the Infrared
Space Observatory or the Spitzer Space Telescope, means that the
second method is more practicable for extragalactic sources, and a
large number of authors have taken this approach -- see for example
\citet{vlo99b} or \citet{gro09} and references therein for mass-loss
derivations for stars in the LMC and SMC. However, these two methods
probe different parts of the circumstellar envelope
\citep[e.g.,][]{kem03}, and often give differing values of total
mass-loss rate.

Mass-loss rates for AGB stars in the Galaxy range from 10$^{-9}$ to
10$^{-4}$\,\msol, with the median being
3$\times$10$^{-7}$\,\msol\ \citep{olo08b,sch01}. A typical mass-loss
rate for a carbon star with a rich circumstellar chemistry such as
IRC+10216 is $\approx$10$^{-5}$\,\msol\ \citep[e.g.,][]{woo03,men01}.
There seems to be no differentiation in mass-loss rate based on
chemistry (M-, C- and S-stars) in the Galactic sample of
\citet{olo08b} (see Fig.~2). There is a weak dependence on metallicity
for O-rich stars compared to C-rich stars, with the difference
appearing to be due to the driving mechanism of the wind
\citep[e.g.][]{hof07}, and the opacity of the dust grains within it
\citep{woi06}.


In the halo of our galaxy, where the metallicity is lower than that in
the plane, a sample of 16 carbon-rich AGB stars have mass-loss rates
of $\sim$4$\times$10$^{-6}$\,\msol, to within a factor of three
\citep{mau08}. Similar, although smaller, rates are estimated by
\citet{gro97} for two of these halo stars. In the metal-rich -- but
still sub-solar -- Sagittarius dwarf spheroidal galaxy, six carbon
stars are detected by \citet{lag10}, who estimate mass-loss rates on
the order of 10$^{-6}$\,\msol.

In the LMC and SMC, carbon stars have similar mass-loss rates to those
in the Galaxy. \citet{tan97} found that dusty carbon-rich LMC stars
have mass-loss rates less than 10$^{-5}$\,\msol; \citet*{lei08} found
that the brightest Magellanic carbon stars which made up their
(biased) sample have mass-loss rates in a narrow range around
10$^{-6}$\,\msol; \citet{vlo03} found a particularly high
mass-loss-rate carbon star, LI-LMC~1813, which has a mass-loss rate of
nearly 4$\times$10$^{-5}$\,\msol. Since this star is in a cluster, the
birth mass and metallicity could be determined, allowing for an
estimate of the dust-to-gas ratio (see~\S\ref{sec:g2d}). Various carbon
stars have been observed in the SMC
\citep*[e.g.,][]{vlo99a,mat05,vlo08} and rates are found to be similar
to those in the LMC \citep[see also][]{vlo00,vlo06}.

Since in this work we are only considering carbon stars, for our
fiducial models in all three metallicity regimes we will assume
$\dot{M}$=3$\times$10$^{-5}$\,\msol. This is toward the high end of
the observed range of mass-loss rates, but typical of stars where the
strong infrared C$_2$H$_2$ features are seen most clearly, and of the
well-known Galactic carbon star IRC+10216.

\subsection{Gas-to-dust ratio}
\label{sec:g2d}

In the early 1990s \citet*{hab94} postulated that the dust-to-gas
ratio would depend on metallicity, and this was refined by
\citet{vlo00} who determined observationally that it had an
approximately linear dependence. Dust production at low metallicity is
limited by the availability of heavy metals (e.g., titanium) to form
the condensation seeds for dust formation \citep{vlo08}.

There have been several attempts to determine gas-to-dust ratios in
the diffuse ISM of the MCs from H{\sc i} and infrared maps. At higher
densities this becomes almost impossible. Moreover, it is not clear
how the gas-to-dust ratio in the diffuse ISM relates to that in
circumstellar environments. In the ISM of the LMC, the dust-to-gas
ratio is approximately a quarter that of the solar value, and this has
been established for several decades
\citep[e.g.,][]{vge70,koo82,cla85} and confirmed more recently with
the Herschel and Spitzer telescopes \citep[e.g.,][]{mei10,gor03}. In
the SMC the ISM dust-to-gas ratio is about a tenth that of solar
\citep{vdb68,vge70,leq82,bou85,bot04,gor09}. Polycyclic Aromatic
Hydrocarbons behave differently, but are also depleted at low
metallicity \citep{san12}.

In the circumstellar envelopes of AGB stars, the dust-to-gas ratio is
harder to determine, especially without a good measure of the gas
mass. This ratio also depends on chemistry: in \textit{oxygen-rich}
envelopes the dependence is roughly linear \citep{mar04}. In
carbon-rich envelopes there is some decline with metallicity, but
possibly slightly shallower than linear \citep{vlo00}. Estimates
derived from molecular band strengths \citep[e.g.,][]{slo06} can be
misleading \citep{vlo08}. In lieu of firm determinations, authors have
assumed various values: \citet{vlo03} assume a gas-to-dust value of
300--500 for the carbon star LI-LMC~1813; \citet{lei08} assume 100 for
the Galaxy, 200 for the LMC and 500 for the SMC. We adopt 100, 200 and
500, respectively, in line with the metallicity, for the fiducial
models.

\subsection{Wind expansion velocity}

The expansion velocity of a circumstellar envelope, that is, the
terminal velocity once the wind has undergone acceleration in and
close to the dust formation zone, depends on the dust-to-gas ratio,
which in turn depends on the metallicity. There is also a lesser
reliance upon the luminosity of the star, such that
$v_\mathrm{exp}\propto\Psi^{\nicefrac{1}{2}}L^{\nicefrac{1}{4}}$,
where $\Psi$\ is the dust-to-gas ratio, and $L$\ the luminosity
\citep{hab94,eli01,mar04}. Thus expansion velocity is expected to be
lower in the Magellanic Clouds, for a given luminosity.

There is also a link between expansion velocity and the mass-loss
mechanism, since pulsation-driven winds (which are associated with low
mass-loss rates) show low expansion velocities. Stars experiencing a
superwind, where the mass-loss driver is radiation pressure on dust
grains, generally show larger expansion velocities
\citep[e.g.,][]{win00}.

In general, expansion velocities in the LMC have been measured to be
fairly small compared to the Galaxy. For all chemistries, Galactic
expansion velocities cover a wide range, from low \citep[e.g.,
  1.5--22.5\,km\,s$^{-1}$;][]{olo08b} to high \citep[e.g.,
  4.3--35.4\,km\,s$^{-1}$;][]{lou93} and even higher velocities are
expected theoretically \citep*[e.g., up to
  60\,km\,s$^{-1}$;][]{mat10}.  For Galactic carbon stars undergoing a
superwind mass-loss, typical expansion velocities range from
13--22\,km\,s$^{-1}$\ \citep[e.g.,][]{woo03}. In the LMC,
\citet{vlo03} derived $v_\mathrm{exp}$=9.5\,km\,s$^{-1}$\ from
modelling the carbon star LI-LMC~1813. In a sample of {\it
  oxygen-rich} circumstellar envelopes in the LMC, \citet{mar04}
measured the expansion velocity of a number of OH masers and obtained
results mostly in the region 8--17\,km\,s$^{-1}$. Circumstellar OH
masers are thought to trace the terminal velocity of the wind, rather
than H$_2$O masers, for example, which trace the acceleration zone.
\citet*{woo98} suggested that stars with higher expansion velocities
have higher metallicities. More recently, \citet[][and references
  therein]{vlo00,mar04} and \citet{sch07a} have suggested that
expansion velocities increase with evolution on the AGB. Expansion
velocities for stars in the SMC have not been determined. It should be
noted that gas velocities and dust velocities are not necessarily
well coupled, and drift velocities can be substantial at low mass-loss
rates \citep[e.g.,][]{jon01}.

One can also look to the Galactic Halo as a low-metallicity
environment. Expansion velocities have been determined for several
Halo stars: \citet{gro97} measured an expansion velocity of
3.2\,km\,s$^{-1}$ from a $^{12}$CO J=2--1 line, which is exceptionally
low for a carbon star. This value was confirmed by \citet{lag10}, who
observed the $^{12}$CO J=3--2 line. They also calculated a low (dust)
mass-loss rate for this object. \citet{lag10} observed two other
carbon stars in the Halo, and determined expansion velocities of 6.5
and 8.5\,km\,s$^{-1}$. Three other carbon-rich objects in their sample
were thought to be in the Galactic Halo, but instead reside within the
Galaxy's more metal-rich thick disc. Expansion velocities for these
stars were measured as 11.5--16.5\,km\,s$^{-1}$. Thus it seems clear
that for carbon stars, at least, expansion velocity decreases with
decreasing metallicity. This is consistent with theoretical models of
AGB outflows \citep{wac08}, where low expansion velocities are due to
the formation of less dust, which in turn implies a less efficient
acceleration of the circumstellar material.

For the purposes of our fiducial models, we have adopted
$v_\mathrm{exp}$=20, 10 and 5\,km\,s$^{-1}$\ for the Galaxy, LMC and
SMC respectively.

\subsection{UV radiation field}
\label{sec:isrf}


\citet{ber08} were able to establish that the interstellar radiation
field in the LMC is in general stronger than that in the Milky
Way. They find that it varies from 0.8\,G$_0$\ in diffuse regions to
3.5\,G$_0$\ in molecular regions, with an average value of
$\sim$2\,G$_0$. We use this value in our modelling. Such a galactic
variation in field strength is also seen in the Galaxy, where the
strength at the inner molecular ring can be up to five times that in
the solar neighbourhood \citep{pal07}. For the SMC, recent work by
\citet[][Fig.~10]{san10} has shown that the ISRF has an average value
of $\sim$4\,G$_0$, although it can be as high as 10--30\,G$_0$\ in
regions of star formation.

\begin{table}
\caption{Physical parameters of the fiducial models}
\label{tab:physparams}
\begin{tabular}{@{}lccc}
\hline
Parameter & MW & LMC & SMC \\
\hline
$\dot{M}$\ (\msol) & 3$\times$10$^{-5}$ & 3$\times$10$^{-5}$ & 3$\times$10$^{-5}$ \\
1/$\Psi$\                        & 100 & 200 & 500 \\
$v_\mathrm{exp}$\ (km\,s$^{-1}$)  & 20 & 10 & 5 \\
$G$\ (G$_0$)                     & 1 & 2 & 4 \\
\hline
\end{tabular}
\newline$\dot{M}$\ is the mass-loss rate, $\Psi$\ the dust-to-gas
ratio, $v_\mathrm{exp}$\ the envelope expansion velocity, and $G$\ the
interstellar UV field.
\end{table}

\subsection{Caveats}
\label{sec:caveats}

Although we are presenting here three fiducial, and exploratory,
  models, not intended to represent any particular source, it is worth
  noting sources of uncertainties in our models, and how they may
  affect chemistry and predicted line intensities. Many of the
  physical parameters above have mainly an effect on the radius at
  which UV photons will dominate the chemistry of the
  envelope. Decreasing the mass-loss rate, increasing the expansion
  velocity of the wind or ISRF strength, or decreasing the dust-to-gas
  ratio (i.e., reducing the dust content) will have the effect of
  allowing greater penetration for UV photons, meaning that the
  envelope chemistry transitions to a photochemistry (rather than an
  ion-molecule or neutral-neutral chemistry) closer to the star. In
  practice, this means that parent molecules are destroyed more
  effectively, daughter species are produced and then destroyed in a
  narrower shell, and the distribution of species becomes more
  condensed. Since abundant parent species are found in the more dense
  parts of the envelope, where column densities are high, any changes
  to this threshold radius will have a minimal effect on the emission
  intensity from parents.


\section{Chemical modelling}

Our chemical network is based on that of \citet{cor09}, augmented with
silicon chemistry from the UMIST Database for Astrochemistry 2006
\citep[UDfA;][]{woo07}. Reactions with Si-bearing species are included
only for those species already present in the network of
\citet{cor09}. The resulting network contains 481 chemical species,
including atoms, neutral molecules, cations and anions, linked by 6174
reactions, which is considerably larger than the most recent release
of UDfA, Rate06. Species range in size from single atoms (H, He, C,
Mg, N, O, S, Si) to long hydrocarbon chains (e.g., C$_{23}$,
C$_{23}$H$_2$).

We follow a parcel of gas as it passes from the inner edge of the
circumstellar envelope outwards at the terminal wind velocity. Since
we assume a constant mass-loss process, this procedure is time
independent, and we obtain a snapshot of the chemical structure of the
circumstellar envelope during the AGB phase. Chemical rate equations
are solved using the Gear method for stiff differential equations
\citep{gea71}, resulting in fractional abundances for all species at
each radial gridpoint.

\section{Results}

\begin{figure}
\includegraphics[width=84mm]{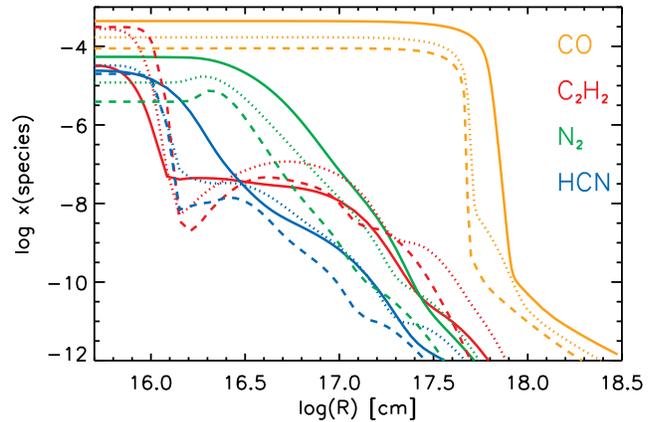}
\caption{Fractional abundances of TE parent species. The solid line
  indicates the Galactic model, the dotted line is the LMC model, and
  the dashed line is the the SMC model.}
\label{fig:full-StabPar}
\end{figure}

\begin{figure}
\includegraphics[width=84mm]{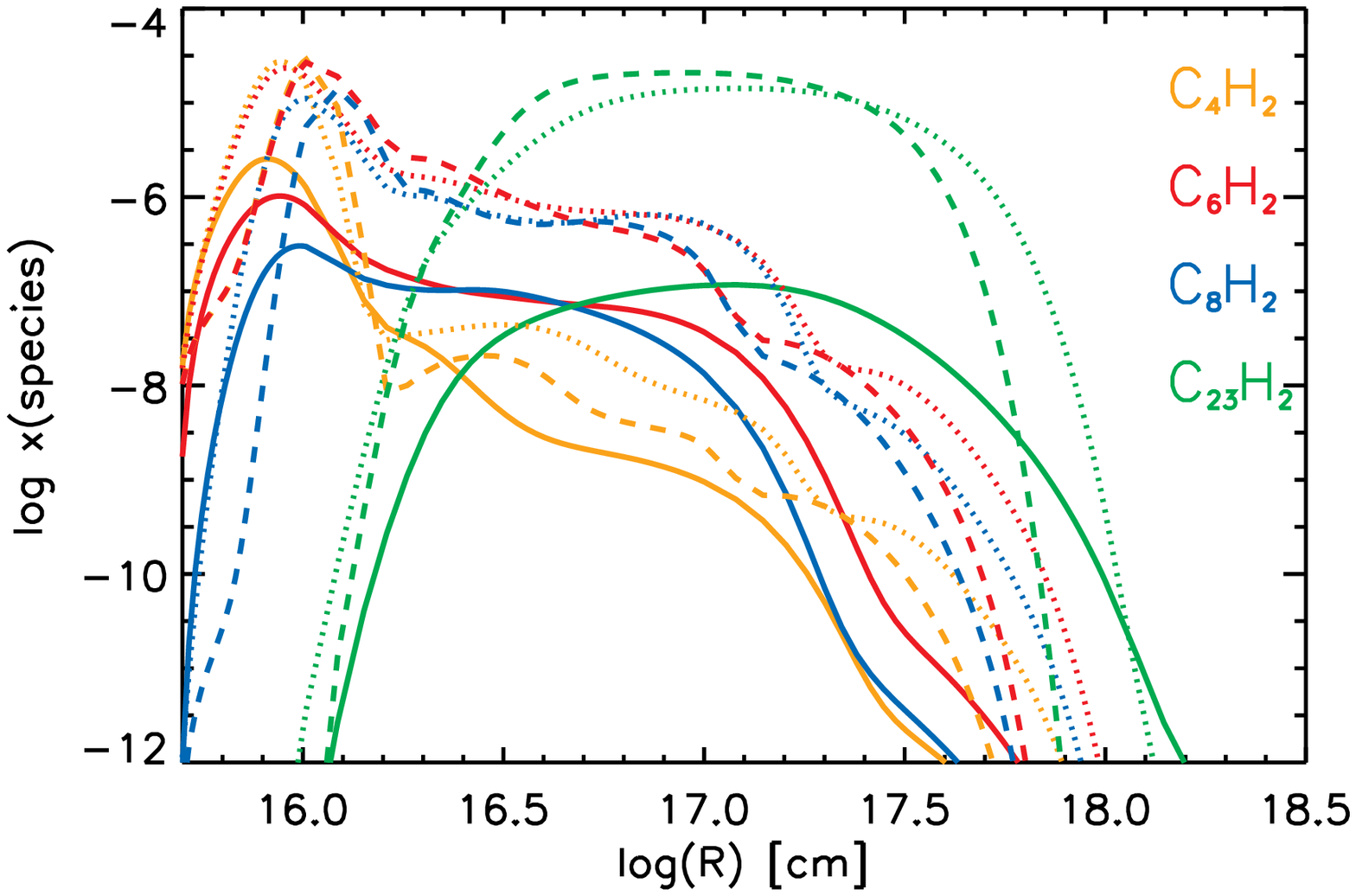}
\caption{Fractional abundances of the polyyne species. Line characteristics are
  as in Fig.~\ref{fig:full-StabPar}.}
\label{fig:full-Polyyne}
\end{figure}

\begin{figure}
\includegraphics[width=84mm]{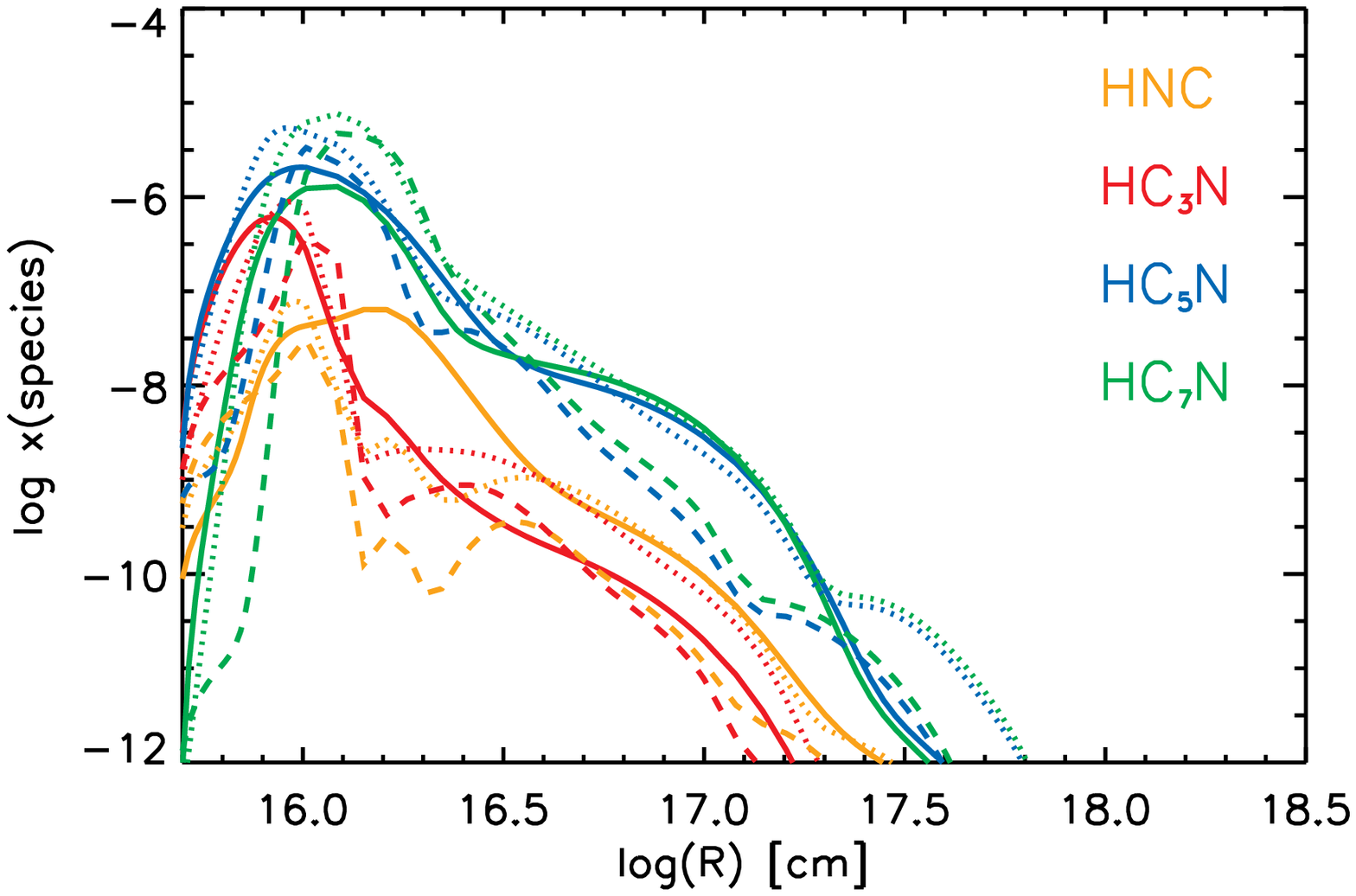}
\caption{Fractional abundances of HNC and cyanopolyyne species. Line characteristics are as in Fig.~\ref{fig:full-StabPar}.}
\label{fig:full-CyPoly}
\end{figure}

\begin{figure*}
\includegraphics[width=165mm]{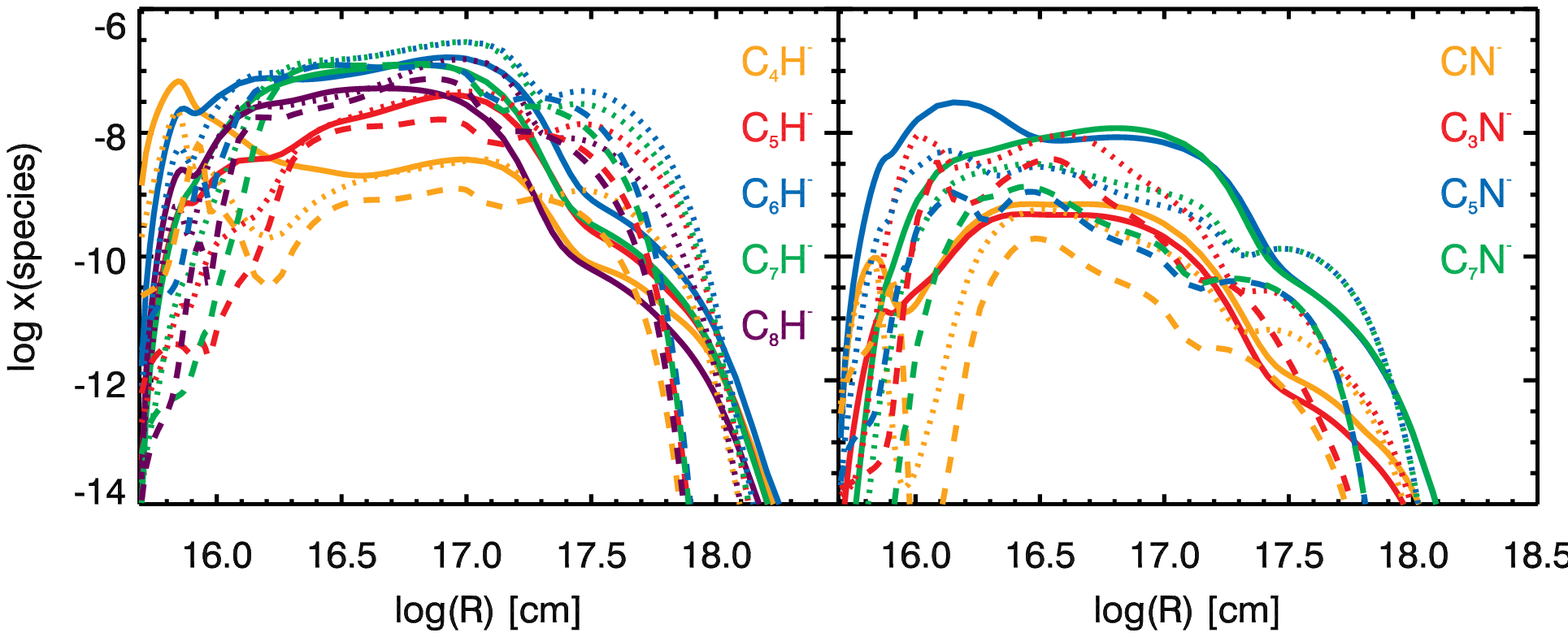}
\caption{Fractional abundances of the anion species. Line characteristics are
  as in Fig.~\ref{fig:full-StabPar}.}
\label{fig:full-Anion2}
\end{figure*}

\begin{figure}
\includegraphics[width=84mm]{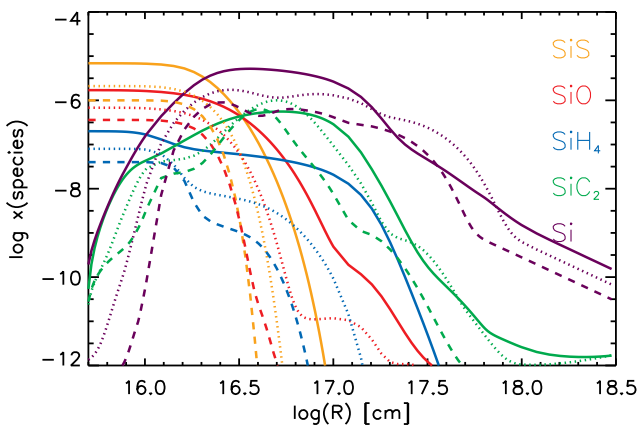}
\caption{Fractional abundances of the silicon-bearing species. Line characteristics are
  as in Fig.~\ref{fig:full-StabPar}.}
\label{fig:full-Sil}
\end{figure}

Radial abundance profiles for various species are plotted in
Figs.~\ref{fig:full-StabPar}--\ref{fig:full-Sil}, and we discuss
different families of molecules as follows.


\subsection{Carbon monoxide, polyynes and cyanopolyynes}

Carbon monoxide (CO) in Galactic carbon stars is the most abundant
molecule after H$_2$, and the most readily observable in the
submillimetre regime, given its high abundance and significant dipole
moment. However, it is clear from the calculations that at lower
metallicity (even at just half solar metallicity in the LMC),
acetylene (C$_2$H$_2$) becomes the dominant carbon-bearing molecule
for $R\lesssim$10$^{16}$\,cm (Fig.~\ref{fig:full-StabPar}). In terms
of column density through the envelope, N(C$_2$H$_2$)$\gtrsim$N(CO)
for LMC metallicity and is three times greater at SMC metallicity. The
CO abundance is lower in LMC and SMC carbon stars, and this reduces
the efficiency of self-shielding, meaning that the extent of the CO
envelope is smaller in comparison.


The predominance of acetylene at lower metallicity means that the
hydrocarbon chemistry is also more developed in such environments
(Fig.~\ref{fig:full-Polyyne}). Acetylene and the ethynyl (C$_2$H)
radical are the basis of much of the carbon-chain growth that occurs
in carbon-rich circumstellar environments \citep{mil00}, and so larger
species such as triacetylene (C$_6$H$_2$) experience a large boost in
production. In the LMC and SMC models, N(C$_6$H$_2$) =
3.5--3.9$\times$10$^{17}$\,cm$^{-2}$, whereas in the Galactic model
the column density is 30 times smaller. The peak abundance of
C$_{23}$H$_2$, the largest polyyne in the model, is comparable for the
two lower metallicity regimes, but $\sim$100 times smaller for the
Galactic model. This increase in large hydrocarbons also applies to
the cyanopolyyne chains (HC$_3$N, HC$_5$N, etc.)  despite the
reduction in elemental nitrogen abundance
(Fig.~\ref{fig:full-CyPoly}). This would indicate that as metallicity
decreases, nitrogen is preferentially sequestered in HCN and the
cyanopolyynes rather than in other nitrogen-bearing
species. Table~\ref{tab:nitrogenrepos}, which gives an overview of the
nitrogen repositories in the model by comparing column densities
through the envelope, shows that this is the case. The abundance of
HC$_x$N (odd $x\ge3$) species increases from Galactic metallicity to
LMC metallicity but decreases at SMC metallicities, following the
changes in initial fractional abundance of HCN
(Table~\ref{tab:initabunds}).

The distribution of the cyanopolyyne chains is very different to that
seen in \citet[][Fig.~6]{mil00}; however it matches very well with the
smooth density distribution model of \citet{cor09}. The difference
arises from the inclusion of an increased photodissociation rate for
HC$_3$N, resulting in a substantial reduction of the HC$_3$N/HC$_5$N
ratio.

\begin{table}
  \caption{Main nitrogen repositories at differing metallicities.}
\label{tab:nitrogenrepos}
\begin{tabular}{@{}lccc}
\hline
Species & MW & LMC & SMC \\
\hline
N$_2$           & 71\% & 37\% & 25\% \\
HCN             & 14\% & 42\% & 58\% \\
N               & 12\% & 12\% & 12\% \\
HC$_{\Sigma(3\ldots11)}$N & \phantom{1}1\% & \phantom{1}6\% & \phantom{1}3\% \\
NH$_3$          & \phantom{1}1\%  & \phantom{1}1\% & \phantom{1}1\% \\
CN              & $<$1\% &  $<$1\% &  $<$1\% \\
\hline
\end{tabular}
\end{table}

\subsection{Anions}

\begin{table*}
  \caption{Column densities and ion-neutral ratios for anionic species in IRC+10216.}
\label{tab:anions}
\begin{tabular}{@{}lccccc}
\hline
Species & Observed column & Neutral-ion & Reference &  Calculated column & Calculated \\
        & density (cm$^{-2}$) & ratio   &           & density (cm$^{-2}$) & neutral-ion ratio \\
\hline
CN$^-$     & 8.0$\times$10$^{12}$ & 400  & \citet{agu10} & 3.0$\times$10$^{12}$ & $\sim$8\,000 \\
C$_3$N$^-$ & 1.6$\times$10$^{12}$ & 200  & \citet{tha08} & 1.6$\times$10$^{12}$ & $\sim$240\\
C$_5$N$^-$ & 3.4$\times$10$^{12}$ & 2    & \citet{cer08} & 1.3$\times$10$^{14}$ & 6 \\
C$_2$H$^-$ & $<$7$\times$10$^{10}$ & $>$70\,000 & \citet{agu10} & 8.1$\times$10$^{8\phantom{1}}$ & $\sim$10$^8$ \\
C$_4$H$^-$ & 7.1$\times$10$^{11}$ & 4\,200 & \citet{cer07} & 7.2$\times$10$^{14}$ & 17 \\
C$_6$H$^-$ & 4.0$\times$10$^{12}$ & 16--100 & \citet{mcc06,cer07} & 8.2$\times$10$^{14}$ & 4 \\
C$_8$H$^-$ & 2.4$\times$10$^{12}$ & 3--4 & \citet{rem07,kaw07} & 2.1$\times$10$^{14}$ & 5\\
\hline
\end{tabular}
\end{table*}

Several anions have now been detected in Galactic carbon star
IRC+10216: C$_6$H$^-$\ was identified by \citet{mcc06}, and confirmed
by \citet{cer07}, who also reported the initial detection of
C$_4$H$^-$. C$_8$H$^-$\ was detected by \citet{rem07} and
\citet{kaw07}. The anions of the cyanopolyyne radicals C$_3$N and
C$_5$N were discovered by \citet{tha08} and \citet{cer08},
respectively. Most recently \citet{agu10} detected the smallest
molecular anion to date, CN$^-$. A summary of observed column
densities and estimates of neutral-to-anion ratios is shown in
Table~\ref{tab:anions}. The anion chemistry is discussed in detail by
\citet{cor09}.

In general, the radiative electron attachment rate of these carbon
chains increases with length \citep{her08}, however, the abundance of
the neutrals from which the anions are created peaks with C$_6$H, and
hence the most abundant anions in these models are C$_6$H$^-$\ and
C$_7$H$^-$ (Fig.~\ref{fig:full-Anion2}).  We can compare the observed
column densities and ratios for these anions with those calculated in
our Galactic model (also shown in Table~\ref{tab:anions}). Despite not
specifically modelling IRC+10216, the agreement between model and
observation is reasonably good for the smaller hydrocarbons. For
column densities of the larger hydrocarbon anions, C$_4$H$^-$\ and
larger, the agreement is not as close, with the model overproducing by
factors of 100-1\,000. The abundances of these hydrocarbon anions are
particularly sensitive to the initial abundance of acetylene, as has
been discussed by \citet{rem07}. They show that a reduction in the
initial abundance of acetylene by a factor of 5--10 brings about a
much better overall agreement. The C$_2$H$^-$\ anion has not yet been
detected, and we predict a column density of $\sim$10$^9$\,cm$^{-2}$,
several orders of magnitude below that of the detected anions. Thus,
unless we have underestimated the radiative electron attachment rate
of C$_2$H, C$_2$H$^-$\ is not likely to be detectable. \citet{agu10}
have determined an upper limit for this species in IRC+10216, of
$<$0.0014\%\ of the C$_2$H column (Table~\ref{tab:anions}). Similarly,
C$_{10}$H$^-$\ has not been detected, but has a similar ion-neutral
ratio to C$_6$H$^-$\ and C$_8$H$^-$, and a column density which is
about a quarter of that of C$_8$H$^-$, according to the model. Also
abundant in the envelope are negatively-charged carbon chains,
C$_{4\ldots9}^-$, which have column densities of
8$\times$10$^{13}$--1$\times$10$^{15}$\,cm$^{-2}$.

The distribution of hydrocarbon anions (Fig.~\ref{fig:full-Anion2}) is
similar for all members of the family from C$_4$H$^-$\ to
C$_8$H$^-$. Similar distributions are also found for
C$_{1,3,5,7}$N$^-$. At the inner peak of its distribution, at log($R$)
= 15.85, CN$^-$\ is formed via electron attachment to MgNC, even in
the lowest metallicity case. \citet{agu10} find that \mbox{HCN +
  H$^-$\ $\longrightarrow$\ CN$^-$\ + H$_2$} provides a minor
contribution in this region, but we find this reaction
ineffective. The much broader, outer peak of the CN$^-$\ distribution
arises due to the reaction \mbox{$\mathrm{N} + \mathrm{C}_{4\ldots7}^-
  \longrightarrow\ \mathrm{CN}^- + \mathrm{C}_{3\ldots6}$}. In
general, the larger members of the family are all formed by electron
attachment to corresponding neutrals. The distribution of CN$^-$\ fits
well with the observations of \citet{agu10}, and the column density
calculated in the model (3.0$\times$10$^{12}$\,cm$^{-2}$) matches the
observed determination (8.0$\times$10$^{12}$\,cm$^{-2}$) very
closely. Similarly, the column density of
C$_3$N$^-$\ (1.6$\times$10$^{12}$\,cm$^{-2}$) matches that determined
for IRC+10216 by \citet{tha08} exactly. \citet{cer08} derived a rather
low column density for C$_5$N$^-$\ of
3.4$\times$10$^{12}$\,cm$^{-2}$\ (lower than for CN$^-$), but note
that this figure is likely to be an underestimate. We calculate a
column density of $N$(C$_5$N$^-$)$=$1.3$\times$10$^{14}$\,cm$^{-2}$.

At lower metallicity, small anions are a factor of a few lower in
abundance than at Galactic metallicity. This is surprising given that
the corresponding neutrals are more abundant, and the number of free
electrons in the Magellanic envelopes is larger. The larger anions
C$_6$H$^-$ and C$_7$H$^-$ are predicted to be more abundant at LMC
metallicity than Galactic.

\subsection{Silicon chemistry}

Silicon chemistry was not included in the model of \citet{cor09}, but
is included here due to the potentially important reaction:
\begin{equation}\label{eq:sic2form}
\mathrm{C_2H_2} + \mathrm{Si} \longrightarrow \mathrm{SiC_2}\ + \mathrm{H_2},
\end{equation}
which could be a major sink of C$_2$H$_2$\ in the inner regions of the
envelope at low metallicity. However, the model shows that reactions
with other hydrocarbons are the major loss mechanisms for C$_2$H$_2$,
even in the highest metallicity case (i.e., where the abundance of
silicon atoms is highest).

SiC$_2$\ was recently detected by the Herschel Space Observatory in
IRC+10216 \citep{cer10}, where its distribution traces the
dust-formation zone of the star, similar to SiO and SiS emission
\citep{fon08}. However, it also has a significant abundance in the
outer envelope \citep{luc95,gen95}, as we find in our modelling (see
Fig.~\ref{fig:full-Sil}), with the distribution peaking at
$\sim$5$\times$10$^{16}$\,cm and reaching
$x$(SiC$_2$)$=$5$\times$10$^{-7}$. These figures correspond well with
the observations of \citet{luc95} and the chemical model of
\citet{cer10}. The abundance of SiC$_2$\ increases through the
reaction between acetylene and silicon atoms (Eq.~\ref{eq:sic2form}),
and it becomes destroyed outside $R$=10$^{17}$\,cm by reaction with
abundant C$^+$\ ions. The peak abundance of SiC$_2$\ is similar in all
metallicity regimes, and column densities are slightly enhanced in the
LMC and SMC models
(1.9$\times$10$^{15}$:4.2$\times$10$^{15}$:2.6$\times$10$^{15}$;
MW:LMC:SMC). The chemistry is slightly different in the LMC and SMC
models, since the neutralisation of SiC$_2$H$^+$\ to form
SiC$_2$\ dominates over
reaction~(\ref{eq:sic2form}). SiC$_2$H$^+$\ itself is partially formed
from SiC$_2$, via the chain:
\begin{eqnarray}
\nonumber\mathrm{Si^+} + \mathrm{C_2H} &\longrightarrow& \mathrm{SiC_2^+} + \mathrm{H}\\
\nonumber\mathrm{C^+} + \mathrm{SiC_2} &\longrightarrow& \mathrm{SiC_2^+} + \mathrm{C}\\
\nonumber\mathrm{SiC_2^+} + \mathrm{H_2} &\longrightarrow& \mathrm{SiC_2H^+} + \mathrm{H}\\
\nonumber\mathrm{SiC_2H^+} + \mathrm{e^-~(or~anion)} &\longrightarrow& \mathrm{SiC_2} + \mathrm{H}\: \mathrm{(}+ \mathrm{neutral)}
\end{eqnarray}

\begin{table}
  \caption{Main silicon repositories at differing metallicities.}
\label{tab:siliconrepos}
\begin{tabular}{@{}lccc}
\hline
Species & MW & LMC & SMC \\
\hline
SiS           & 70\%   & 65\% & 64\% \\
SiO           & 17\%   & 21\% & 23\% \\
Si            & \phantom{1}6\% & \phantom{1}6\% & \phantom{1}5\% \\
Si$^+$        & \phantom{1}6\% & \phantom{1}4\% & \phantom{1}6\% \\
SiH$_4$       & \phantom{1}2\%  & \phantom{1}2\% & \phantom{1}2\% \\
SiC$_2$       & $<$1\% &  \phantom{1}2\% & \phantom{1}2\%  \\
\hline
\end{tabular}
\end{table}

The major repository for silicon in circumstellar envelopes is SiS
(Table~\ref{tab:siliconrepos}), in line with the high initial
abundances of this molecule (Table~\ref{tab:initabunds}). This is true
in all metallicity regimes tested, although the abundance of SiS drops
slightly with metallicity as sulphur is preferentially incorporated
into CS.

\subsection{The effect of assumptions about physical parameters}

\begin{table*}
  \caption{Calculated column densities (cm$^{-2}$) for species of
    interest.}
\label{tab:coldens}
\begin{tabular}{@{}lcccc|c}
\hline
Species & $N_\mathrm{MW}$ & $N_\mathrm{LMC}$ & $N_\mathrm{SMC}$ & Ratio & Ratio$^\mathrm{fixed}_\mathrm{physics}$ \\
\hline
H$_2$	&	2.88$\times$10$^{22}$	&	5.77$\times$10$^{22}$	&	1.15$\times$10$^{23}$	&	1.00:\phantom{1}2.00:\phantom{1}4.00	&	1.00:\phantom{1}1.00:\phantom{1}1.00	\\

C$_2$H	&	7.88$\times$10$^{16}$	&	7.33$\times$10$^{16}$	&	2.64$\times$10$^{16}$	&	1.00:\phantom{1}0.93:\phantom{1}0.33	&	1.00:\phantom{1}1.71:\phantom{1}1.78	\\
{\bf C$_2$H$^-$}	&	8.12$\times$10$^{8\phantom{0}}$	&	7.48$\times$10$^{8\phantom{0}}$	&	1.10$\times$10$^{9\phantom{0}}$	&	1.00:\phantom{1}0.92:\phantom{1}1.36	&	1.00:\phantom{1}0.94:\phantom{1}1.02	\\
{\bf C$_2$H$_2$}	&	5.28$\times$10$^{17}$	&	1.06$\times$10$^{19}$	&	2.89$\times$10$^{19}$	&	1.00:20.08:54.77	&	1.00:\phantom{1}7.95:\phantom{1}8.76	\\
C$_2$S	&	2.67$\times$10$^{14}$	&	3.43$\times$10$^{14}$	&	1.78$\times$10$^{14}$	&	1.00:\phantom{1}1.28:\phantom{1}0.67	&	1.00:\phantom{1}0.99:\phantom{1}0.50	\\
CH$_3$CN	&	1.65$\times$10$^{13}$	&	1.88$\times$10$^{12}$	&	2.81$\times$10$^{12}$	&	1.00:\phantom{1}0.11:\phantom{1}0.17	&	1.00:\phantom{1}0.18:\phantom{1}0.07	\\
CN	&	2.46$\times$10$^{16}$	&	1.48$\times$10$^{16}$	&	8.84$\times$10$^{15}$	&	1.00:\phantom{1}0.60:\phantom{1}0.36	&	1.00:\phantom{1}0.62:\phantom{1}0.35	\\
CN$^-$	&	3.01$\times$10$^{12}$	&	2.93$\times$10$^{12}$	&	1.46$\times$10$^{12}$	&	1.00:\phantom{1}0.97:\phantom{1}0.48	&	1.00:\phantom{1}0.92:\phantom{1}0.47	\\
CO	&	1.29$\times$10$^{19}$	&	9.93$\times$10$^{18}$	&	1.04$\times$10$^{19}$	&	1.00:\phantom{1}0.77:\phantom{1}0.81	&	1.00:\phantom{1}0.39:\phantom{1}0.20	\\
CS	&	5.66$\times$10$^{16}$	&	7.67$\times$10$^{16}$	&	8.01$\times$10$^{16}$	&	1.00:\phantom{1}1.36:\phantom{1}1.42	&	1.00:\phantom{1}0.65:\phantom{1}0.32	\\
C$_3$H	&	5.92$\times$10$^{14}$	&	1.04$\times$10$^{15}$	&	6.29$\times$10$^{14}$	&	1.00:\phantom{1}1.75:\phantom{1}1.06	&	1.00:\phantom{1}2.01:\phantom{1}1.81	\\
C$_3$H$_2$	&	3.03$\times$10$^{14}$	&	7.37$\times$10$^{14}$	&	8.19$\times$10$^{14}$	&	1.00:\phantom{1}2.43:\phantom{1}2.70	&	1.00:\phantom{1}2.80:\phantom{1}2.53	\\
C$_3$N	&	3.92$\times$10$^{14}$	&	6.08$\times$10$^{14}$	&	1.87$\times$10$^{14}$	&	1.00:\phantom{1}1.55:\phantom{1}0.48	&	1.00:\phantom{1}1.63:\phantom{1}0.95	\\
C$_3$N$^-$	&	1.64$\times$10$^{12}$	&	1.77$\times$10$^{12}$	&	7.93$\times$10$^{11}$	&	1.00:\phantom{1}1.08:\phantom{1}0.48	&	1.00:\phantom{1}1.13:\phantom{1}0.60	\\
C$_3$S	&	1.03$\times$10$^{15}$	&	1.47$\times$10$^{15}$	&	5.87$\times$10$^{14}$	&	1.00:\phantom{1}1.43:\phantom{1}0.57	&	1.00:\phantom{1}1.20:\phantom{1}0.62	\\
C$_4$H	&	1.21$\times$10$^{16}$	&	2.00$\times$10$^{16}$	&	6.57$\times$10$^{15}$	&	1.00:\phantom{1}1.65:\phantom{1}0.54	&	1.00:\phantom{1}3.30:\phantom{1}3.86	\\
C$_4$H$^-$	&	7.18$\times$10$^{14}$	&	3.04$\times$10$^{14}$	&	1.11$\times$10$^{14}$	&	1.00:\phantom{1}0.42:\phantom{1}0.15	&	1.00:\phantom{1}0.99:\phantom{1}0.93	\\
C$_4$H$_2$	&	2.89$\times$10$^{16}$	&	4.35$\times$10$^{17}$	&	3.55$\times$10$^{17}$	&	1.00:15.04:12.26	&	1.00:11.63:12.85	\\
{\bf C$_5$N}	&	7.51$\times$10$^{14}$	&	1.84$\times$10$^{15}$	&	6.28$\times$10$^{14}$	&	1.00:\phantom{1}2.45:\phantom{1}0.84	&	1.00:\phantom{1}2.79:\phantom{1}1.80	\\
C$_5$N$^-$	&	1.33$\times$10$^{14}$	&	4.10$\times$10$^{13}$	&	9.11$\times$10$^{12}$	&	1.00:\phantom{1}0.31:\phantom{1}0.07	&	1.00:\phantom{1}0.52:\phantom{1}0.29	\\
C$_6$H$^-$	&	8.24$\times$10$^{14}$	&	1.84$\times$10$^{15}$	&	1.73$\times$10$^{15}$	&	1.00:\phantom{1}2.23:\phantom{1}2.09	&	1.00:\phantom{1}2.16:\phantom{1}2.31	\\
{\bf C$_6$H$_2$}	&	1.15$\times$10$^{16}$	&	3.86$\times$10$^{17}$	&	3.51$\times$10$^{17}$	&	1.00:33.55:30.50	&	1.00:24.63:29.99	\\
C$_8$H$^-$	&	2.14$\times$10$^{14}$	&	7.35$\times$10$^{14}$	&	8.22$\times$10$^{14}$	&	1.00:\phantom{1}3.44:\phantom{1}3.85	&	1.00:\phantom{1}2.07:\phantom{1}2.13	\\
HCN	&	5.45$\times$10$^{17}$	&	1.44$\times$10$^{18}$	&	1.94$\times$10$^{18}$	&	1.00:\phantom{1}2.64:\phantom{1}3.55	&	1.00:\phantom{1}1.20:\phantom{1}0.71	\\
HC$_3$N	&	6.44$\times$10$^{15}$	&	1.27$\times$10$^{16}$	&	5.26$\times$10$^{15}$	&	1.00:\phantom{1}1.97:\phantom{1}0.82	&	1.00:\phantom{1}1.74:\phantom{1}1.02	\\
{\bf HC$_5$N}	&	1.89$\times$10$^{16}$	&	8.66$\times$10$^{16}$	&	3.97$\times$10$^{16}$	&	1.00:\phantom{1}4.58:\phantom{1}2.09	&	1.00:\phantom{1}3.59:\phantom{1}2.31	\\
HNC	&	2.95$\times$10$^{14}$	&	9.11$\times$10$^{14}$	&	7.46$\times$10$^{14}$	&	1.00:\phantom{1}3.08:\phantom{1}2.53	&	1.00:\phantom{1}4.77:\phantom{1}3.62	\\
MgNC	&	3.92$\times$10$^{15}$	&	3.01$\times$10$^{15}$	&	1.81$\times$10$^{15}$	&	1.00:\phantom{1}0.77:\phantom{1}0.46	&	1.00:\phantom{1}0.39:\phantom{1}0.15	\\
SiC	&	1.10$\times$10$^{14}$	&	5.22$\times$10$^{13}$	&	3.02$\times$10$^{13}$	&	1.00:\phantom{1}0.48:\phantom{1}0.27	&	1.00:\phantom{1}0.43:\phantom{1}0.24	\\
SiC$_2$	&	1.93$\times$10$^{15}$	&	4.17$\times$10$^{15}$	&	2.56$\times$10$^{15}$	&	1.00:\phantom{1}2.16:\phantom{1}1.33	&	1.00:\phantom{1}0.92:\phantom{1}0.47	\\
SiN	&	1.05$\times$10$^{14}$	&	2.48$\times$10$^{13}$	&	1.19$\times$10$^{13}$	&	1.00:\phantom{1}0.24:\phantom{1}0.11	&	1.00:\phantom{1}0.15:\phantom{1}0.04	\\
SiO	&	4.37$\times$10$^{16}$	&	3.49$\times$10$^{16}$	&	3.70$\times$10$^{16}$	&	1.00:\phantom{1}0.80:\phantom{1}0.85	&	1.00:\phantom{1}0.40:\phantom{1}0.21	\\
SiS	&	1.77$\times$10$^{17}$	&	1.08$\times$10$^{17}$	&	1.03$\times$10$^{17}$	&	1.00:\phantom{1}0.61:\phantom{1}0.58	&	1.00:\phantom{1}0.31:\phantom{1}0.15	\\
\hline
\end{tabular}
~\\ List of molecules observed by \citet{he08} in IRC+10216, with the
addition of those in bold face.
\end{table*}

In order to fully understand the chemistry at low metallicity distinct
from the effect of the differing physical conditions, we calculated
three models adopting the physical conditions of the Galactic model
(Table~\ref{tab:physparams}), and using the initial chemical
abundances of the three different metallicity regimes. Hence, any
variation in abundance or column density is solely due to the
chemistry of the circumstellar envelope, or the initial abundances
adopted. For clarity, we will name these the \textquotedblleft fixed
physics\textquotedblright\ models, as opposed to the \textquotedblleft
full physics\textquotedblright\ models, as described previously.

In Table~\ref{tab:coldens} we present the column densities derived
from the full physics models, and their ratios. In the final column we
have listed the ratio of column densities derived from the fixed
physics models. Looking at how the two final columns in the table
change gives us an indication of how the adopted physical conditions
affect the chemistry of the model.

For many of the species metallicity is the dominating factor, with
column densities falling with metallicity despite an increase in the
overall H$_2$\ column density. Many of the nitrogen-bearing species
are included in this group, which makes them good tracers of
metallicity: CN, CH$_3$CN, C$_5$N$^-$. However, there are exceptions
to this: C$_3$N, C$_5$N, C$_3$N$^-$\ and the cyanopolyynes increase in
relative column density at LMC metallicity, but decrease at SMC
metallicity. Species containing heavy metals such as silicon or
magnesium also reflect changes in metallicity, with the exception to
this being SiC$_2$. For SiC$_2$\ the effect of the lowering
metallicity is to produce a decrease in the column density (see final
column of Table~\ref{tab:coldens}). However, column densities in the
full physics models increase for LMC and SMC metallicities,
counteracting the effects due to chemistry alone. Another species with
similar properties is CS, which is produced inherently less in lower
metallicity environments, but receives a boost in production in the
denser environments of the full physics models.

CO is not a very good tracer of metallicity in carbon stars, despite
it being a good tracer of oxygen abundance under the fixed physics
conditions. Observationally, CO emission is often optically thick, and
thus the observed flux intensity is not representative of the
abundance of CO. In addition, in our models column densities of CO in
the full LMC and SMC models are comparable, despite a factor of two
difference in metallicity. CO is a very good tracer of the molecular
envelope of carbon stars, though, and gives us an indication (see
Fig.~\ref{fig:full-StabPar}) that in general, the envelopes of
Magellanic carbon stars will only extend $\sim$70\% of the distance of
similar Galactic carbon stars.

The column densities of hydrocarbons such as C$_2$H$_2$, C$_4$H,
C$_4$H$_2$\ and C$_6$H$_2$\ increase significantly at lower
metallicity, as we have seen. Table~\ref{tab:coldens} shows that much
of this increase is due to the changes in chemistry at lower
metallicity, with the relative proportion of carbon increasing over
oxygen (Table~\ref{tab:COratios}). A significant increase in column
density for these species is also due to the differing physical
conditions found in Magellanic carbon stars. These species are formed
through neutral-neutral reactions, and the higher densities of
Magellanic envelopes means that these reactions proceed more
quickly. This trait is not observed to the same degree in anions of
hydrocarbon species, which are generally more abundant than in
Galactic carbon stars for larger species (C$_6$H$^-$\ and larger), but
only by a factor of a few. This is somewhat surprising, since the
ionisation fraction of the envelope almost doubles from Galactic model
to LMC model to SMC model. Smaller hydrocarbon species (C$_4$H$^-$,
C$_5$H$^-$) are more abundant in Galactic CSEs.

\section{Observables}

With the impending completion of ALMA in 2013, it is interesting to
consider the potential for ALMA to observe molecular line emission
from extragalactic AGB CSEs, focusing on the Magellanic Clouds. We
have estimated line intensities for nearly half a million lines
assuming local thermal equilibrium (LTE) (Sect.~\ref{sec:LTE}), and
selected a few potential candidate lines for further investigation
using a more rigorous non-LTE radiative transfer code
(Sect.~\ref{sec:NLTE}).

\subsection{LTE estimates}
\label{sec:LTE}

\begin{table*}
\begin{minipage}{165mm}
  \caption{Predicted line strengths for carbon stars in the LMC, using Eq.~\ref{eq:lmclteintens}.}
\label{tab:lineintens}
\begin{tabular}{lcclcclcc}
\hline
\multicolumn{3}{c}{Band 3 (89-119\,GHz)}&\multicolumn{3}{c}{Band 6 (211-275\,GHz)}&\multicolumn{3}{c}{Band 7 (275-370\,GHz)}\\
Molecule & Frequency & Peak Flux & Molecule & Frequency & Peak Flux & Molecule & Frequency & Peak Flux \\
         & (GHz)       & (mJy)           &          & (GHz)       & (mJy)           &          & (GHz)       & (mJy) \\
\hline
C$_4$H$_2$ & \phantom{0}89.315 & 0.1 & SiO          & 217.105 & \phantom{0}0.7          & CS           & 293.912 & \phantom{0}0.3 \\
C$_4$H$_2$ & \phantom{0}89.687 & 0.1 & $^{13}$CO    & 220.399 & \phantom{0}1.0          & SiO          & 303.927 & \phantom{0}0.3 \\       
SiC$_2$    & \phantom{0}93.064 & 0.1 & CN           & 226.632 & \phantom{0}0.2          & $^{13}$CO    & 330.588 & \phantom{0}1.7 \\
C$_4$H$_2$ & \phantom{0}97.834 & 0.1 & CN           & 226.660 & \phantom{0}0.6          & CN           & 340.008 & \phantom{0}0.1 \\
CS         & \phantom{0}97.981 & 0.1 & CN           & 226.664 & \phantom{0}0.2          & CN           & 340.020 & \phantom{0}0.1 \\
C$_4$H$_2$ & \phantom{0}98.245 & 0.1 & CN           & 226.679 & \phantom{0}0.2          & CN           & 340.032 & \phantom{0}1.1 \\
C$_4$H$_2$ & \phantom{0}98.655 & 0.1 & CN           & 226.874 & \phantom{0}0.6          & CN           & 340.035 & \phantom{0}0.4 \\
C$_4$H$_2$ & 107.175           & 0.1 & CN           & 226.875 & \phantom{0}0.9          & CN           & 340.035 & \phantom{0}0.7 \\
$^{13}$CO   & 110.201          & 0.1 & CN           & 226.876 & \phantom{0}0.4          & CN           & 340.248 & \phantom{0}1.0 \\
CN         & 113.491           & 0.1 & CN           & 226.887 & \phantom{0}0.1          & CN           & 340.248 & \phantom{0}1.4 \\
CO         & 115.271           & 2.5 & CN           & 226.892 & \phantom{0}0.1          & CN           & 340.249 & \phantom{0}0.8 \\
SiC$_2$    & 115.382           & 0.2 & CO           & 230.538 & $\approx$40  & CS           & 342.883 & \phantom{0}0.1 \\
C$_4$H$_2$ & 116.105           & 0.1 & CS           & 244.936 & \phantom{0}0.5          & CO            & 345.796 & $\approx$40 \\ 
           &                   &     & SiO          & 260.518 & \phantom{0}0.5          & SiO           & 347.331 & \phantom{0}0.1 \\
           &                   &     & C$_2$H       & 262.004 & \phantom{0}1.5          & C$_2$H         & 349.338 & \phantom{0}1.3 \\
           &                   &     & C$_2$H       & 262.006 & \phantom{0}1.1          & C$_2$H         & 349.339 & \phantom{0}1.0 \\ 
           &                   &     & C$_2$H       & 262.065 & \phantom{0}1.1          & C$_2$H         & 349.399 & \phantom{0}1.0 \\ 
           &                   &     & C$_2$H       & 262.067 & \phantom{0}0.7          & C$_2$H         & 349.401 & \phantom{0}0.8 \\
           &                   &     & c-C$_3$H$_2$ & 265.759 & \phantom{0}0.1          & HCN            & 354.505 & 58.9 \\
           &                   &     & HCN          & 265.886 &            66.7         & c-C$_3$H$_2$   & 368.294 & \phantom{0}0.4 \\ 
\hline                                            
\end{tabular}

\medskip                                     
{\bf Band 9 (602-720\,GHz):} Strongest lines include C$_2$H$_3^+$, HCN
and CO, all with intensities on the order of 1\,mJy.
\end{minipage}
\end{table*}


 Following the calculations of \citet{olo08a}, one can write:
\begin{equation}
  S_\mathrm{CO(2\mbox{--}1)}\approx6\left[\frac{\dot{M}}{10^{-6}}\right]^{1.2}\left[\frac{15}{v_\mathrm{e}}\right]^{1.6}\left[\frac{X_\mathrm{CO}}{10^{-3}}\right]^{0.7}\left[\frac{1}{D}\right]^2~\mathrm{Jy},
\end{equation}
where the CO ($J$=2--1) line flux density is given in terms of the
mass-loss rate, expansion velocity, fraction of CO with respect to
H$_2$, and the distance, $D$, to the star. Inserting typical values
for the LMC gives us an estimated flux density of $\approx$0.04
Jy. Similarly, for the SMC, we calculate 
$S_\mathrm{CO(2\mbox{--}1)}\approx$0.09 Jy, since we expect the
density of circumstellar envelopes in the SMC to be higher. Comparable
flux densities are expected for the CO ($J$=3--2) line. Such stars
should be easily detectable within an hour's observations for the full
ALMA array (5$\sigma$\ = 6~mJy at 2~km s$^{-1}$\ resolution). Stars
with properties similar to our LMC stellar characteristics should be
detectable out to $\approx$150\,kpc within one hour, whilst stars
similar to our SMC stellar characteristics could potentially be seen
out to distances of $\approx$200\,kpc within an hour \citep{olo08a}.

For species other than CO, the line flux density can also be estimated
\citep{olo08a}. For LMC carbon stars,
\begin{equation}
  S\approx 1.62\times10^{-9}g_\mathrm{u}A_\mathrm{ul}f_\mathrm{X}R_\mathrm{e}\frac{e^{-E_\mathrm{l}/kT_\mathrm{X}}}{Q(T_\mathrm{X})} \mathrm{Jy}, \label{eq:lmcmols} \label{eq:lmclteintens}
\end{equation}
using our standard parameters, and where $g_\mathrm{u}$,
$A_\mathrm{ul}$\ and $E_\mathrm{l}$\ are the quantum mechanical
degeneracy of the upper energy level of the transition, the Einstein
coefficient for the transition, and the lower energy level,
respectively. $Q(T_\mathrm{X})$\ is the partition function, dependent
on the excitation temperature of the molecule, $T_\mathrm{X}$, which
we assume to be 10 K for all rotational transitions
\citep[cf.][]{woo03}. $R_\mathrm{e}$\ is the radius of the emitting
region for each molecule, taken from the molecular distributions in
the chemical model. Similarly for the SMC,
\begin{equation}
  S\approx 3.72\times10^{-9}g_\mathrm{u}A_\mathrm{ul}f_\mathrm{X}R_\mathrm{e}\frac{e^{-E_\mathrm{l}/kT_\mathrm{X}}}{Q(T_\mathrm{X})} \mathrm{Jy}. \label{eq:smcmols} \label{eq:smclteintens}
\end{equation}
Estimated flux densities for molecular lines stronger than 0.1\,mJy
are given in Table \ref{tab:lineintens}. The peak line intensities
calculated from Eqs.~\ref{eq:lmclteintens} and ~\ref{eq:smclteintens}
generally underestimate line strengths by a factor of $\approx$5--50
in comparison to those from the non-LTE model (\S\ref{sec:NLTE}). The
exception to this is lines of HCN, which are over-estimated by
$\approx$9--16 times in comparison to the NLTE estimates. Our
assumption of $T_\mathrm{X}=10$\,K for HCN lines is probably a
significant underestimate for a species which is only abundant in the
inner regions of the circumstellar envelope. Adopting
$T_\mathrm{X}=75$\,K reduces $S_\mathrm{HCN(3\mbox{--}2)}\approx$27
mJy, approximately halving the estimate for $T_\mathrm{X}=10$\,K.

\subsection{Non-LTE estimates}
\label{sec:NLTE}

\begin{figure*}
\includegraphics[width=84mm]{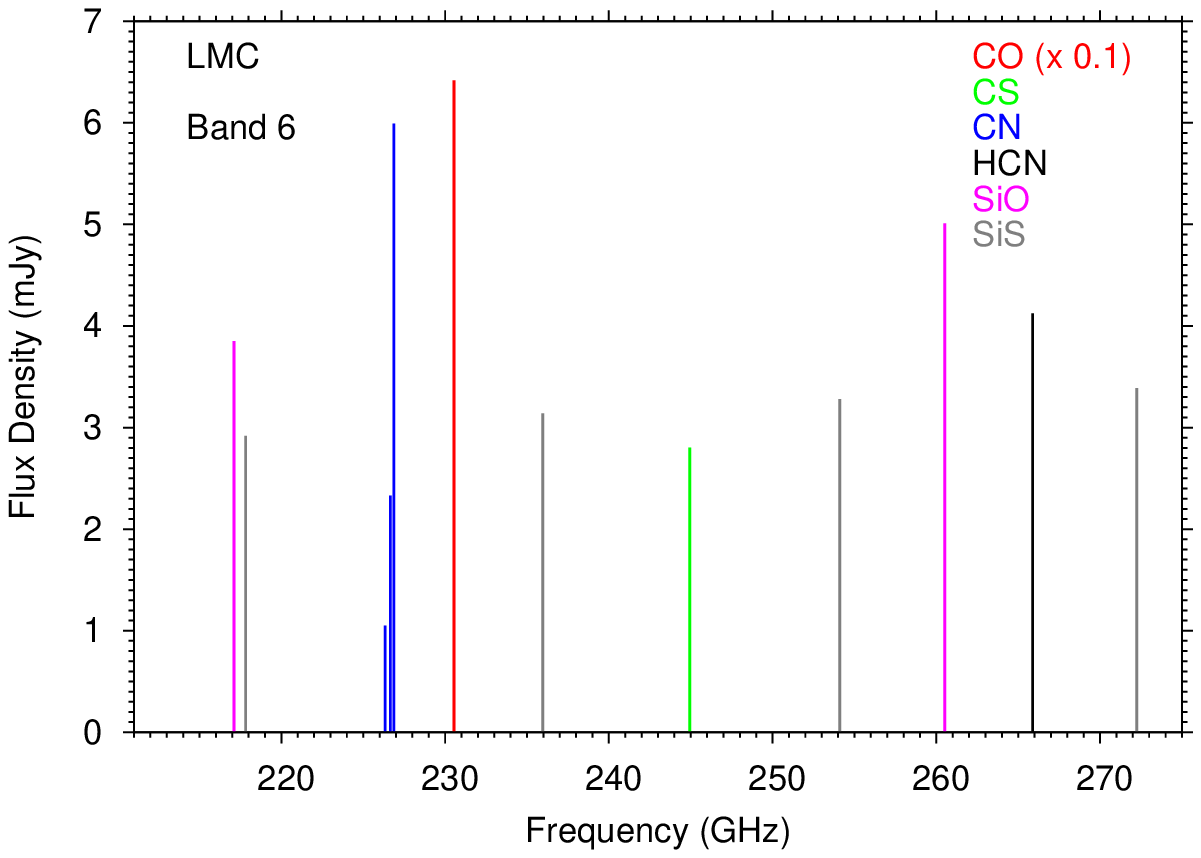}\includegraphics[width=84mm]{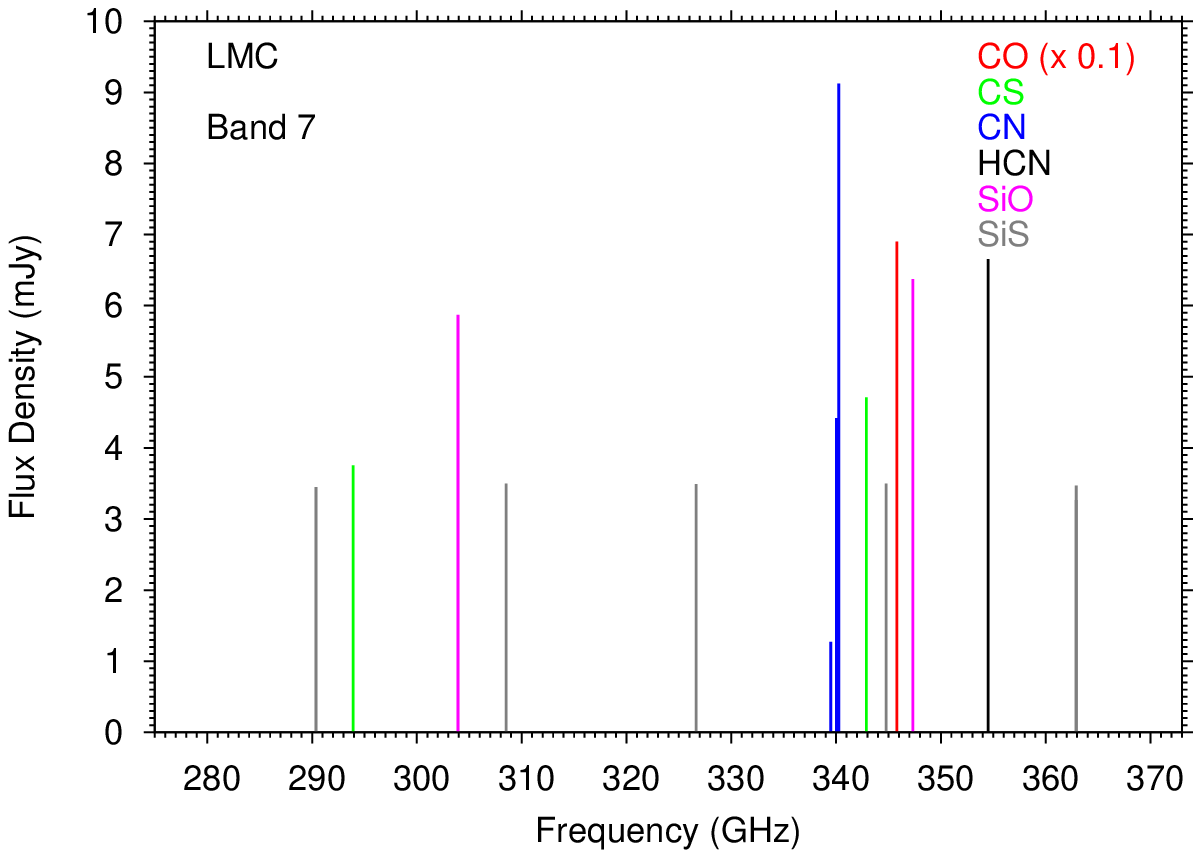}
\caption{Synthetic spectra of an LMC carbon star for ALMA Band 6 and Band 7.}
\label{fig:LMC_6}
\end{figure*}
\begin{figure*}
\includegraphics[width=84mm]{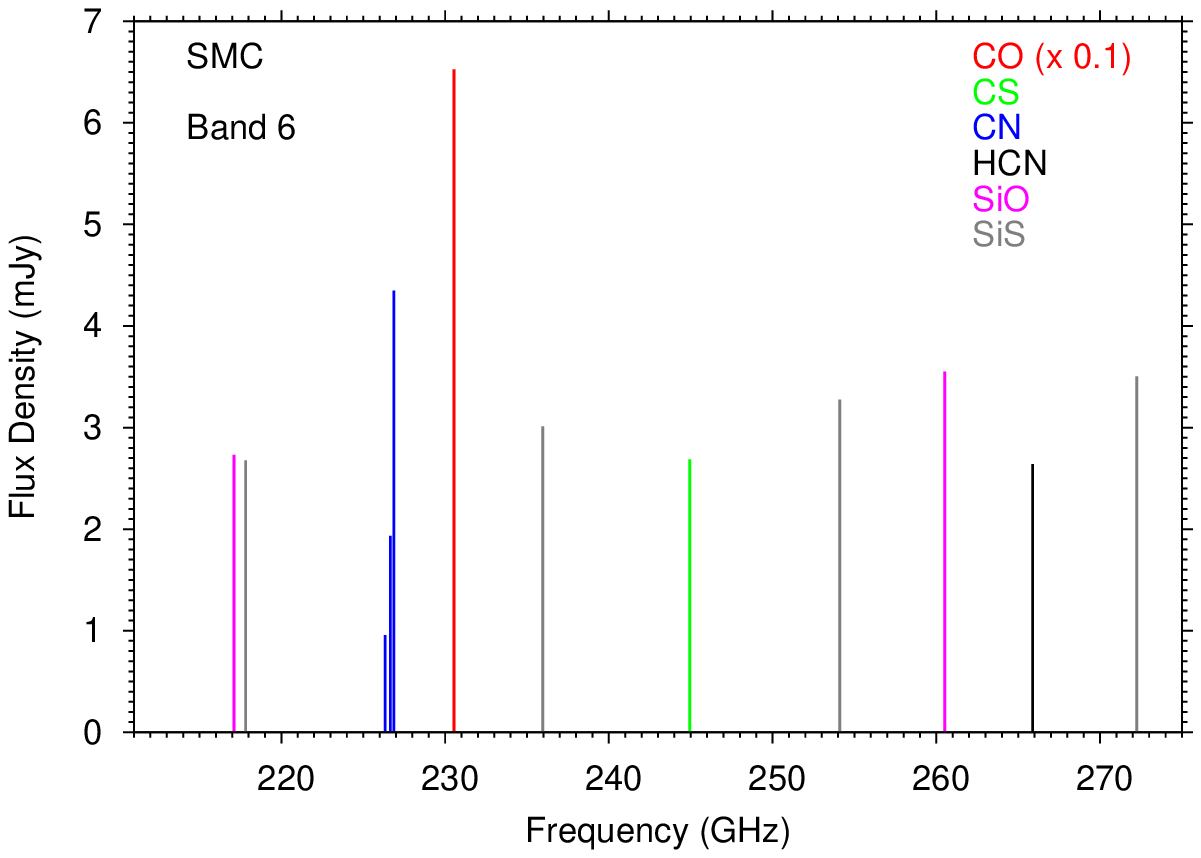}\includegraphics[width=84mm]{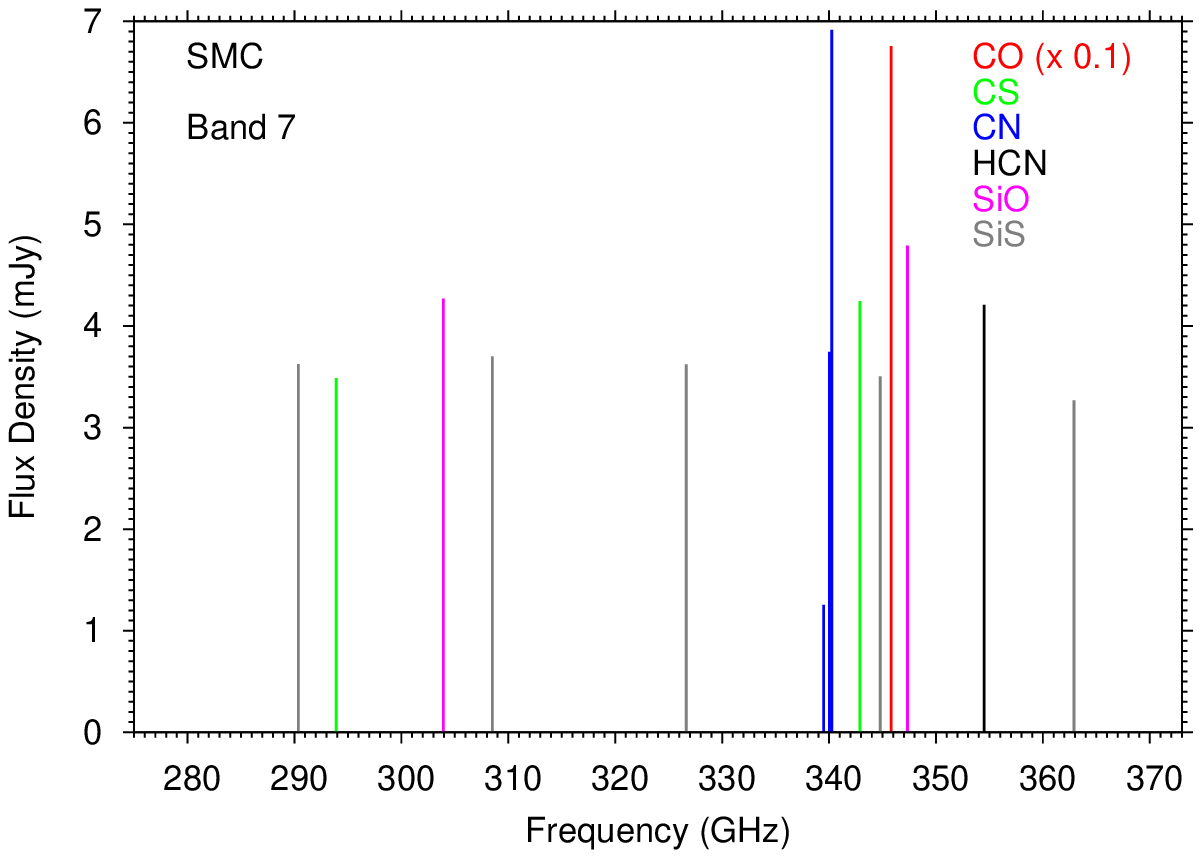}
\caption{Synthetic spectra of an SMC carbon star for ALMA Band 6 and Band 7.}
\label{fig:SMC_6}
\end{figure*}

\begin{table*}
\begin{minipage}{165mm}
\caption{Estimated observation times for ALMA Full Science, using the ALMA Sensitivity Calculator.}
\label{tab:ALMA_obs}
\begin{tabular}{c c c c c c c c c}
\hline
 & & & \multicolumn{3}{c}{LMC} & \multicolumn{3}{c}{SMC} \\
Molecule & Transition & Frequency   & Peak Flux       & S/N & Integration Time & Peak Flux       & S/N & Integration Time \\
         &            & (GHz)       & (mJy)           &     &             & (mJy)           &     & \\
\hline
CO       & J = 1--0   & 115.271         & 19.2            & 10  & 50.4 min         & 19.8            & 10  &  47.2 min \\
         & J = 2--1   & 230.528         & 64.2            & 10  &  1.3 min         & 65.3            & 10  &   1.3 min \\
         & J = 3--2   & 345.796         & 69.0            & 10  &  2.8 min         & 67.5            & 10  &   3.0 min \\
         & J = 4--3   & 461.041         & 65.4            & 10  &  2.6 hr          & 57.8            & 10  &   3.9 hr  \\ 
         & J = 6--5   & 691.473         & 64.2            &  5  &  1.4 hr          & 52.9            &  5  &   2.5 hr  \\
HCN      & J = 3--2   & 265.886         & \phantom{0}4.1  &  5  &  1.6 hr          & \phantom{0}2.6  &  5  &   4.2 hr  \\
         & J = 4--3   & 354.505         & \phantom{0}6.7  &  5  &  1.7 hr          & \phantom{0}4.2  &  5  &   4.5 hr  \\
CS       & J = 5--4   & 244.936         & \phantom{0}2.8  &  5  &  2.6 hr          & \phantom{0}2.7  &  5  &   2.6 hr  \\
         & J = 6--5   & 293.912         & \phantom{0}3.8  &  5  &  2.1 hr          & \phantom{0}3.5  &  5  &   2.8 hr  \\
         & J = 7--6   & 342.883         & \phantom{0}4.7  &  5  &  2.7 hr          & \phantom{0}4.3  &  5  &   2.8 hr  \\
SiO      & J = 5--4   & 217.105         & \phantom{0}3.9  &  5  &  1.4 hr          & \phantom{0}2.7  &  5  &   3.6 hr  \\        
         & J = 6--5   & 260.518         & \phantom{0}5.0  &  5  &  1.0 hr          & \phantom{0}3.6  &  5  &   2.0 hr  \\
         & J = 7--6   & 303.927         & \phantom{0}5.9  &  5  &  1.0 hr          & \phantom{0}4.3  &  5  &   1.8 hr  \\
         & J = 8--7   & 347.331         & \phantom{0}6.4  &  5  &  1.3 hr          & \phantom{0}4.8  &  5  &   2.4 hr  \\
SiS      & J = 12--11 & 217.817         & \phantom{0}2.9  &  5  &  2.5 hr          & \phantom{0}2.7  &  5  &   3.6 hr  \\
         & J = 13--12 & 235.961         & \phantom{0}3.1  &  5  &  2.7 hr          & \phantom{0}3.0  &  5  &   2.8 hr  \\
         & J = 14--13 & 254.103         & \phantom{0}3.3  &  5  &  2.0 hr          & \phantom{0}3.3  &  5  &   2.0 hr  \\
         & J = 15--14 & 272.243         & \phantom{0}3.4  &  5  &  2.1 hr          & \phantom{0}3.5  &  5  &   2.2 hr  \\
         & J = 16--15 & 290.380         & \phantom{0}3.5  &  5  &  2.6 hr          & \phantom{0}3.6  &  5  &   2.7 hr  \\
         & J = 17--16 & 308.516         & \phantom{0}3.5  &  5  &  3.2 hr          & \phantom{0}3.7  &  5  &   3.3 hr  \\
CN       & J = 2--1    & 226.333         & \phantom{0}2.3  &  5  &  3.6 hr         & \phantom{0}1.9  &  5  &   5.7 hr  \\
         & J = 2--1    & 226.659         & \phantom{0}6.0  &  5  & 37.6 min        & \phantom{0}4.4  &  5  &   1.1 hr  \\
         & J = 3--2    & 340.031         & \phantom{0}4.4  &  5  &  2.7 hr         & \phantom{0}3.7  &  5  &   4.6 hr  \\
         & J = 3--2    & 340.249         & \phantom{0}9.1  &  5  & 39.9 min        & \phantom{0}6.9  &  5  &   1.1 hr  \\
\hline
\end{tabular}

\medskip
In our integration time estimates, we assume an array size of 50 antennae and a spectral resolution of 2.0\,km/s.
\end{minipage}
\end{table*}

In addition to our simple estimates for the expected line flux density
described above, we have used the one-dimensional non-LTE Monte Carlo
radiative transfer code, RATRAN \citep{hog00}, to estimate the line
profiles and line strengths of molecular line emission from evolved
carbon stars in the LMC and SMC.  In our simple estimates (see
Eqs.~\ref{eq:lmcmols} and \ref{eq:smcmols}), we assume LTE and an
excitation temperature of 10\,K for all rotational transitions.  In
reality, the envelopes of AGB stars have a radially-dependent
temperature and density structure \citep[see e.g.,][]{jur81} which
limits the accuracy of calculations made with these assumptions.  The
temperature varies from $\sim$1\,000\,K in the inner envelope to
$\sim$10\,K in the outer envelope.  Similarly, the density structure
exhibits a $R^{-2}$\ behaviour so that in the outer envelope the
density is much lower than the critical density of rotational
transitions ($A_{ul}$/$\Sigma_{i} C_{ui}$, where $i$\ indicates all
energy levels lower than the upper level, $u$).  We used molecular
data from the Leiden Atomic and Molecular Database
(LAMDA\footnote{\url{http://www.strw.leidenuniv.nl/~moldata/}}) which
has tabulated energy levels, Einstein A coefficients and collisional
rates for many molecules with transitions in the (sub-)mm region of the
electromagnetic spectrum \citep{sch05}.

In our radiative transfer calculations, as in the physical and
chemical modelling, we treat the stellar envelope as spherically
symmetric with a constant outward stellar wind velocity of
10\,km\,s$^{-1}$\ for the LMC and 5\,km\,s$^{-1}$\ for the SMC.  Since
our line profiles are broadened to a width $\gtrsim$20\,km\,s$^{-1}$\
and $\gtrsim$10\,km\,s$^{-1}$, respectively, we use a spectral
resolution of 1\,km\,s$^{-1}$\ for both sources.  We assume a distance
to source of 50\,kpc for the LMC and 66\,kpc for the SMC.  We
generated line profiles and line strengths for rotational transitions
of CO, CS, CN, HCN, SiO and SiS which fall into the expected spectral
range of ALMA \textquotedblleft Full Science\textquotedblright\
operations ($\approx$30\,GHz to $\approx$950\,GHz).  These are the
molecules which our calculations suggested may be sufficiently
abundant to possess emission strong enough to be observable and for
which collisional rates are available.  Using the ALMA sensitivity
calculator\footnote{\url{almascience.eso.org/call-for-proposals/sensivity-calculator}},
we determined those line transitions which may be observable with ALMA
Full Science within a realistic observing time.  Our results are
listed in Table~\ref{tab:ALMA_obs}.  We determine that molecular
transitions in Bands 6 and 7 make particularly good targets. Assuming
a spectral resolution of 2\,km\,s$^{-1}$\ and an array size of 50
antennae, we find the observation times for LMC and SMC molecular
transitions range from a few minutes for the CO $J$=2--1 and $J$=3--2
transitions at 230 and 345\,GHz to several hours for transitions of
HCN, CS, SiO, SiS and CN which also fall into Bands 6 and 7. In
addition, the CO $J$=1--0 transition, in Band 3, may also be
observable to high signal-to-noise within one hour in evolved stars
within both galaxies. We note here that our time calculations are
likely upper estimates since the ALMA full array may eventually
consist of up to 66 antennae in total. Also, the online ALMA
sensitivity calculator overestimates the predicted observing times
relative to the more accurate calculator available using the ALMA
Observing Tool, again, indicating that our predicted times are likely
upper estimates.  In Figs.~\ref{fig:LMC_6}--\ref{fig:SMC_6} we present
the expected line spectra due to emission from the listed molecules
from a carbon star in both the LMC and SMC, for ALMA bands 6 and 7.
Line profiles of each line transition are also provided in online-only
material (Figs.~9--32). Reducing the mass-loss rate of the models by a
factor of three reduces the line strengths in Table~\ref{tab:ALMA_obs}
by a similar factor. Thus probing the molecular inventory of carbon
stars with low mass-loss rates in the MCs will be challenging.

Looking firstly at the predicted line spectra for the LMC
(Fig.~\ref{fig:LMC_6}), we see that the strongest transitions are due
to CO, CN, HCN and SiO. Multiple transitions of CN, SiO, CS and SiS
are available within a single band which would allow determination of
the temperature and density of the emitting gas, enabling some
constraints on the physical conditions in the envelope. For the SMC
(Fig.~\ref{fig:SMC_6}), we see a similar
pattern, however, we now find that transitions of SiS are comparable
in strength with those of SiO and HCN. Comparing the two metallicity
regimes, we see differences in the line strength ratios between CO/CN
and SiO/SiS, with these ratios generally decreasing with
metallicity. This follows, given the decreased amount of oxygen
available in the SMC relative to the LMC. We also see the line
strength ratios of CN/HCN slightly increase with metallicity, likely
related to the increased strength of the ISRF in the SMC. Our
calculations demonstrate that there are observable tracers of
metallicity and physical conditions in extragalactic carbon stars
which ALMA will allow us to probe.

\section{Discussion}

This investigation was prompted by the observation of very common and
very deep molecular absorptions due to C$_2$H$_2$\ in the mid-infrared
spectra (5--38\,$\mu$m) of LMC carbon stars \citep[e.g.,][for a number
of examples]{woo11} compared to Galactic carbon stars. This occurs
particularly in the extreme carbon stars, which are losing mass at
very high rates. Of course, the fact that these deep features exist
has been established for several years: \citet{spe06} analysed the
deepest C$_2$H$_2$\ absorption feature observed to date; \citet{zij06}
noted that acetylene bands were stronger at low metallicity;
\citet{lei08} measured equivalent widths of molecular lines;
\citet{slo06} and \citet{lag07} detected acetylene absorptions in SMC
carbon stars that appeared deeper than in corresponding Galactic and
LMC carbon stars.  There have also been studies in the
acetylene-dominated 3.1/3.8\,$\mu$m bands which show similar results:
\citet{vlo99a}, \citet{mat02,mat05}, \citet{vlo06}. \citet{vlo08}
performed a 3\,$\mu$m band study, finding that acetylene absorption
features in LMC and SMC carbon stars were equivalent in depth.  The
origin of the 3\,$\mu$m acetylene band is a mixture of photospheric
and circumstellar origin \citep{vlo06}, whereas the 13.7\,$\mu$m
absorption seen in Spitzer spectra is predominantly circumstellar
\citep{mat06}.  We have shown via thermal equilibrium calculations
that \textquotedblleft photospheric\textquotedblright\ acetylene
increases in abundance as the metallicity is lowered
(Fig.~\ref{fig:te}), qualitatively matching what we observe at
3\,$\mu$m. We have also shown via radially-dependent non-equilibrium
calculations that indeed acetylene is more abundant in carbon-rich
circumstellar envelopes at lower metallicity
(Fig.~\ref{fig:full-StabPar}). Initial investigations into carbon
stars in other low(er)-metallicity galaxies show that acetylene
features are also very strong \citep[e.g.,][]{slo09} Unfortunately we
do not yet have sufficiently good quality infrared data to perform
quantitative analyses.

Related to the study of acetylene is the study of hydrogen cyanide. It
absorbs in the 3.1\,$\mu$m band with acetylene, and also forms part of
the 13.7\,$\mu$m band, absorbing at $\sim$14\,$\mu$m. There is also a
band at 7\,$\mu$m, close to the acetylene band at 7.5\,$\mu$m. Its
contribution to these bands has been detected in a number of
Magellanic carbon stars, but not nearly so prevalently as acetylene
\citep[e.g.,][]{mat05,zij06,vlo06}. In some objects the absorptions
are strong \citep{mat02}, whereas in other objects, absorptions are
very weak \citep{mats08}. Both the thermal equilibrium model and the
circumstellar envelope model predict high abundances of HCN at low
metallicity, despite the lower abundance of elemental nitrogen. This
would imply that both the 3\,$\mu$m and 13.7\,$\mu$m bands should show
ample evidence of HCN absorption. Our high abundances may be an effect
of our choice of 3\,M$_\odot$\ nucleosynthesis models
(Sect.~\ref{sec:initabunds}) since the nitrogen abundance in AGB stars
is dependent on (initial) stellar mass, such that stars with high
luminosity (i.e., high initial mass) should have abundant HCN
\citep{mat05}. The variation in feature strength seen in the infrared
may reflect varying initial stellar mass in the observed samples. The
impact of different choices of stellar mass should be looked into in
future models. Scaling initial abundances up or down uniformly
generally results in a similar scaling of CSE abundances; however,
changes in stellar mass are unlikely to produce such a uniform
scaling, but would result in abundance enhancements in some elements
over others.

Emission from two large circumstellar molecules was observed recently
in Magellanic objects. The fullerenes C$_{60}$\ and C$_{70}$\ have
been detected in a decade of planetary nebulae (PNe; \citealt{gar11};
\citealt{zha11} provides a summary of all fullerene detections) in the
LMC, as well as several Galactic objects, including (proto-)planetary
nebulae and reflection nebulae. The Magellanic clouds are presumably
rich in fullerenes, whether they form from smaller hydrocarbons
coagulating or the destruction of hydrogenated amorphous carbon dust
\citep[e.g.,][]{gar11}. In Galactic nebulae fullerenes potentially
take up one percent of the elemental carbon abundance
\citep{cam10,sel10}; as Table~\ref{tab:COratios} shows, there is
significantly more carbon available for fullerene formation in
Magellanic objects. On average, \citeauthor{gar11}'s sample of
Magellanic carbon-rich PNe contains 60\%\ more C$_{60}$\ than the
Galactic PN, Tc-1. As a representative of large molecules,
C$_{23}$H$_2$\ is two orders of magnitude more abundant in Magellanic
carbon stars than in Galactic (Fig.~\ref{fig:full-Polyyne}) in our
models. As a representative of cyclic molecules, benzene is $\sim$200
times more abundant in Magellanic carbon star envelopes than Galactic;
its column density is $N$(C$_6$H$_6$)$\sim$10$^{-5}N$(C$_2$H$_2$).

These four molecules currently make up the extent of our knowledge of
gas-phase species in the envelopes of carbon stars in the Magellanic
Clouds. Given the derived line intensities and simulated spectra from
Table~\ref{tab:lineintens} and Figs.~\ref{fig:LMC_6}--\ref{fig:SMC_6},
ALMA will be able to detect a handful of molecules in a reasonable
amount of time: detections of CO rotational lines will enable us to
derive mass-loss rates and envelope expansion velocities; detections
of other molecules will enable us to verify our chemical models, and
to improve our understanding of circumstellar carbon chemistry at low
metallicity.

\section{Summary}

Driven by the recent interest in deep molecular absorption features in
Spitzer Space Telescope spectra of evolved carbon stars in the
Magellanic Clouds, we have conducted the first investigation into the
circumstellar chemistry of carbon stars in three metallicity
environments: that of the Galaxy (solar metallicity), the LMC (half
solar metallicity) and the SMC (one fifth solar metallicity). The
general trend is that the abundances of hydrocarbons are greatly
enhanced at low metallicity, so much so that the larger members of
hydrocarbon families are more abundant than smaller members (e.g.,
C$_6$H$_2$\ and C$_8$H$_2$\ are more abundant than C$_4$H$_2$). At
just half of solar metallicity, acetylene becomes more abundant in the
circumstellar envelope than carbon monoxide, the most abundant
molecule apart from H$_2$\ in Galactic CSEs. This is indicative of a
suppressed oxygen chemistry at lower metallicity: other oxygen-bearing
species are also less abundant. Nitrogen chemistry is also suppressed,
except when it is incorporated into hydrocarbons (e.g.,
cyanopolyynes). This means that the main nitrogen repository in the
envelope shifts from N$_2$\ in Galactic stars to HCN in Magellanic
stars. A similar trend is seen for heavy metals (e.g., Si), where
SiC$_2$\ behaves like a hydrocarbon, but other silicon-bearing
molecules are rarer.

Physically, the lower expansion velocities of Magellanic CSEs mean
that the envelopes are more dense. This implies that neutral-neutral
reactions in the inner envelope are more rapid, but the stronger
radiation fields in the LMC and SMC cause the molecular envelopes to
be lesser in extent than in Galactic CSEs. The ionisation fraction
increases as metallicity drops, and the main charge carrier moves from
a heavy metal (Mg$^+$) in Galactic envelopes to C$^+$.

We have compared the results from the Galactic model to the carbon
star IRC+10216, and achieve reasonably good agreement in terms of
molecular abundances and distributions. We discuss the qualitative
results that have been obtained from infrared observations of carbon
stars in the Magellanic Clouds, and also achieve good agreement,
although constraints are few. Finally, we use our models to predict
line intensities and images as seen by ALMA. ALMA is expected to
improve our understanding of carbon stars in the Magellanic Clouds,
and lines of CO, HCN, CS, CN, SiO and SiS should be detectable within
3 hours of observation in LMC carbon stars. Lines of C$_2$H and
$^{13}$CO are predicted to be somewhat weaker, but may also be
observable.

\section*{Acknowledgments}

The authors would like to thank Jacco van Loon for useful comments on
parts of the manuscript. We have made use of the Cologne Database for
Molecular Spectroscopy \citep{mul05,mul01}. We would like to thank the
referee, N.~Mauron, for detailed comments which have resulted in an
improved manuscript.

\bsp

\label{lastpage}

\end{document}